\documentclass[prd,letterpaper,showpacs,groupedaddress,superscriptaddress,nofootinbib,floatfix,preprintnumbers,tightenlines,11pt]{revtex4-1}
\usepackage{hyperref}
\usepackage{amssymb,amsmath,graphicx,color}
\usepackage{bm}
\usepackage[tight]{subfigure}

\def\si{^1 \hskip -0.03in S _0}
\def\siii{^3 \hskip -0.025in S _1}

\def\pislash{{\pi\hskip-0.55em /}}

\def\L1Abar{\tilde{l}_{1,A}}

\def\L{{\Lambda}}

\def\cC{{\mathcal C}}

\def\cO{{\mathcal O}}

\def\cW{{\mathcal W}}
\def\cM{{\mathcal M}}

\def\tnubb{$2\nu\beta\beta$}
\def\znubb{$0\nu\beta\beta$}
\def\eqref#1{{(\ref{#1})}}

\begin{document}

\preprint{INT-PUB-17-002}
\preprint{MIT-CTP-4871}

\begin{figure}[!t]
 \vskip -1.1cm \leftline{
 	\includegraphics[width=3.0 cm]{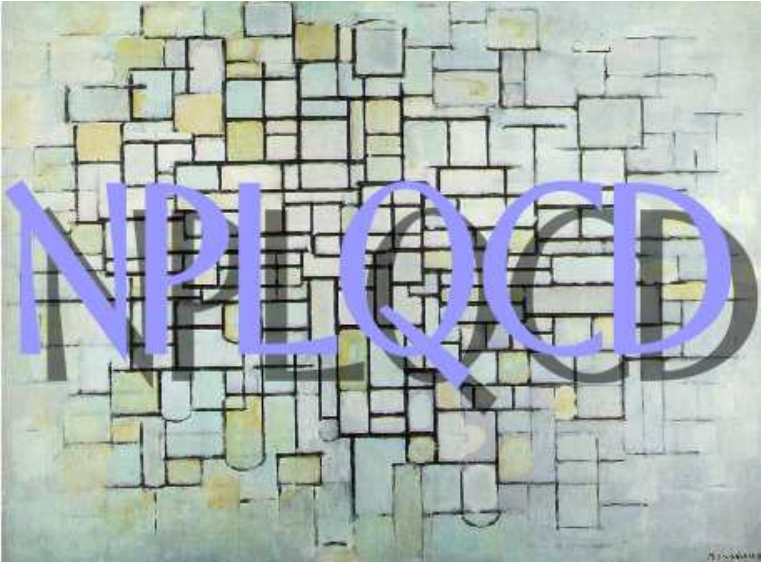}} \vskip
 -0.5cm
\end{figure}

\title{
Double-$\beta$ Decay Matrix Elements
from Lattice Quantum Chromodynamics
}

 \author{Brian C. Tiburzi} 
\affiliation{ Department of Physics, The City College of New York, New York, NY 10031, USA }
\affiliation{Graduate School and University Center, The City University of New York, New York, NY 10016, USA }

\author{Michael L. Wagman} 
\affiliation{Department of Physics,
	University of Washington, Box 351560, Seattle, WA 98195, USA}
	 \affiliation{Institute for Nuclear Theory, University of Washington, Seattle, WA 98195-1550, USA}

\author{Frank Winter}
\affiliation{Jefferson Laboratory, 12000 Jefferson Avenue, 
	Newport News, VA 23606, USA}

\author{Emmanuel~Chang}
\affiliation{Institute for Nuclear Theory, University of Washington, Seattle, WA 98195-1550, USA}

 \author{Zohreh Davoudi} \affiliation{
 	Center for Theoretical Physics, 
 	Massachusetts Institute of Technology, 
 	Cambridge, MA 02139, USA}

 \author{William Detmold} \affiliation{
 	Center for Theoretical Physics, 
 	Massachusetts Institute of Technology, 
 	Cambridge, MA 02139, USA}
	 
 \author{Kostas~Orginos}
 \affiliation{Department of Physics, College of William and Mary, Williamsburg,
 	VA 23187-8795, USA}
 \affiliation{Jefferson Laboratory, 12000 Jefferson Avenue, 
 	Newport News, VA 23606, USA}
 
 \author{Martin J. Savage}
 \affiliation{Institute for Nuclear Theory, University of Washington, Seattle, WA 98195-1550, USA}
 
 \author{Phiala E. Shanahan } \affiliation{
 	Center for Theoretical Physics, 
 	Massachusetts Institute of Technology, 
 	Cambridge, MA 02139, USA} 

\collaboration{NPLQCD Collaboration}

\date{\today}

\pacs{11.15.Ha, 
   12.38.Gc, 
   12.38.-t, 
   21.30.Fe, 
   13.15.+g, 
   23.40.Bw, 
   23.40.-s. 
   }

\begin{abstract}
A lattice quantum chromodynamics (LQCD) calculation of the nuclear matrix element relevant to the $nn\to ppee\overline{\nu}_e\overline{\nu}_e$ transition is described in detail, expanding on the results presented in Ref.~\cite{Shanahan:2017bgi}. This matrix element, which involves two insertions of the weak axial current, is an important input for phenomenological determinations of double-$\beta$ decay rates of nuclei. From this exploratory study, performed using unphysical values of the quark masses, the long-distance deuteron-pole contribution to the matrix element is separated from shorter-distance hadronic contributions. This polarizability, which is only accessible in double-weak processes,  cannot be constrained from single-$\beta$ decay of nuclei, and is found to be smaller than the long-distance contributions in this calculation, but non-negligible. In this work, technical aspects of the LQCD calculations, and of the relevant formalism in the pionless effective field theory, are described. Further calculations of the isotensor axial polarizability, in particular near and at the physical values of the light-quark masses, are required for precise determinations of both two-neutrino and neutrinoless double-$\beta$ decay rates in heavy nuclei. 
\end{abstract}

\maketitle

\section{Introduction}

The second-order weak double-$\beta$ ($\beta\beta$)
decays of nuclei admit important tests of the fundamental symmetries of nature and are probes of the 
Standard Model (SM) and physics beyond it.
The two-neutrino $\beta\beta$-decay mode ($2\nu\beta\beta$), in which the final-state electrons are 
accompanied by two anti-neutrinos, is the rarest SM process that has been measured~\cite{Barabash:2010ie}, and provides crucial tests of our understanding of weak interactions in nuclei. 
Measured $2\nu\beta\beta$-decay rates are benchmark quantities that nuclear many-body calculations 
must reproduce in order for the more complex calculations of neutrinoless $\beta\beta$-decay ($0\nu\beta\beta$)
rates to be considered reliable~\cite{Engel:2016xgb}. 
The $0\nu\beta\beta$-decay mode
can occur only if lepton number is not conserved in nature.
One possible scenario is that a light virtual Majorana neutrino mediates the $\beta\beta$ decay. In this case, $0\nu\beta\beta$-decay rates would be sensitive to the absolute mass scale of neutrinos and could shed light on the neutrino-mass hierarchy~\cite{Mohapatra:2005wg}. 
$0\nu\beta\beta$ decay has not been observed, but it is the primary motivation for a range of current and planned experiments, with at least two orders-of-magnitude improvement in sensitivity expected in the near future~\cite{Avignone:2007fu,Bilenky:2012qi,Dell'Oro:2016dbc}.
Given the significant discovery potential of future $\beta\beta$-decay experiments, it is timely to improve the theoretical understanding of these processes by facilitating the connection of
phenomenological calculations of $\beta\beta$-decay rates to the SM.

Current predictions of nuclear $\beta\beta$-decay rates show significant variation, and their uncertainties are not 
well quantified~\cite{Engel:2016xgb}.
The nuclei that can undergo $\beta\beta$  decay remain too complex for the current {\it ab initio} methods, and there is considerable model dependence in the predictions of the more phenomenological many-body methods that can be applied. 
Moreover, \znubb\ decays may receive contributions from beyond the Standard Model (BSM) physics above the electroweak scale that 
result in short-distance $\Delta L=2$, $\Delta I=2$ operators at hadronic scales (where $L$ and $I$ denote lepton number and isospin, respectively)~\cite{Savage:1998yh, Prezeau:2003xn,Graesser:2016bpz,Cirigliano:2017ymo}.
Additionally, in the light Majorana-neutrino scenario, long-distance second-order weak-current processes are important. These latter contributions are typically modeled using the ``closure approximation'', and other simplifications, whose validity remains to be tested~\cite{Engel:2016xgb}. 
In \tnubb\ decays, the dominant sources of uncertainty are from missing many-body correlations in the nuclear wavefunctions 
and from omitted, or poorly constrained, few-body contributions to the weak currents. 
Reducing these uncertainties is a critical and challenging goal for the nuclear-theory community.

Future planned experiments will likely reduce the uncertainties in \tnubb\ decay rates and thereby better constrain their theoretical description. However, \znubb\ decay-rate calculations with fully quantified uncertainties  require inputs that are not accessible from measurements of processes other than \znubb\ decays and  currently require theoretical inputs.
In light of recent progress in quantitative studies of the
properties of light nuclei from the underlying strong interactions using lattice quantum chromodynamics (LQCD)~\cite{Beane:2012vq,Beane:2013kca,Beane:2014ora,Beane:2015yha,Detmold:2015daa,Chang:2015qxa,Savage:2016kon}, 
it is timely to explore the potential impact of similar SM calculations of nuclear $\beta\beta$ decay.
While nuclear systems that undergo $\beta\beta$ decay are beyond the reach of foreseeable LQCD calculations, computations of the underlying $\beta\beta$-decay matrix elements for small nuclear systems are feasible, as the current work demonstrates. 
Consequently, a promising approach to improving the reliability of $\beta\beta$-decay predictions is to constrain the few-nucleon inputs to 
{\it ab initio} many-body calculations using LQCD studies of the same systems. 
With results from sufficiently precise calculations as input, 
the matching of few and many-body methods, including effective field theories (EFTs), onto the underlying SM interactions
will reduce the uncertainties implicit in many-body approaches, in principle enabling these approaches to provide reliable predictions for \tnubb\ and \znubb\ decay rates.

The symmetries of QCD provide a means to improve some of the uncertainties in SM inputs to $\beta\beta$-decay calculations. Chiral symmetry has been used to relate the $\Delta I={3\over 2}$ amplitude for $K\rightarrow\pi\pi$ to  
pionic matrix elements of a class of short-distance operators inducing \znubb\ decay in Ref.~\cite{Savage:1998yh}, with a more comprehensive study presented recently in Ref.~\cite{Cirigliano:2017ymo} that constrains a larger class of operators. Furthermore, a first attempt to address short-distance $\Delta I=2$ contributions to the 
$\pi^+\to \pi^-$ transition using LQCD is underway, see Ref.~\cite{Nicholson:2016byl} for a preliminary report.\footnote{From an EFT perspective, the effects of induced local operators at the nuclear scale are recovered from both local multi-nucleon operators and through interactions with pions that are exchanged between nucleons as discussed in~\cite{Prezeau:2003xn}. 
There, it is argued that the pionic contribution is dominant, 
although the (Weinberg) power-counting scheme used therein is known to be inconsistent in the $\si$ channel~\cite{Kaplan:1998we, Beane:2001bc}.
} The long-distance second-order weak contributions can also be addressed using LQCD, 
although a number of technical challenges related to double insertions of the operators must be overcome. Recent work by the RBC-UKQCD collaboration~\cite{Christ:2012se,Christ:2015pwa,Christ:2016eae,Christ:2016mmq,Bai:2017fkh} has demonstrated that long-range contributions to the $K_L-K_S$ mass difference, as well the rare kaon-decay matrix elements, can be constrained using LQCD calculations. 
Encouraged by this development, the current work focuses on second-order weak matrix elements in the two-nucleon system.

This work presents the full details of the first LQCD calculation of the forward matrix element of the $I=2$, $I_3=2$ component of the 
time-ordered product of two axial-vector currents in the $\si$ two-nucleon system. 
A synopsis of these results and a discussion of their potential impact on $\beta\beta$-decay phenomenology have been presented in Ref.~\cite{Shanahan:2017bgi}.
Calculations are performed at the $SU(3)$ flavor-symmetric point with degenerate up, down and strange quark 
masses corresponding to a pion mass of $m_\pi\sim806~\tt{MeV}$. Uniform background fields have been successfully implemented in LQCD calculations \cite{Fucito:1982ff,Martinelli:1982cb,Bernard:1982yu} to extract 
magnetic moments and electromagnetic polarizabilities of 
hadrons~\cite{Lee:2005dq,Detmold:2006vu,Detmold:2009dx,Detmold:2010ts,Parreno:2016fwu} 
and nuclei \cite{Beane:2014ora,Chang:2015qxa,Detmold:2015daa}, the magnetic transition amplitude for the $n p \to d \gamma$ process~\cite{Beane:2015yha}, and the axial charge of the proton~\cite{Chambers:2014qaa,Chambers:2015bka}, while generalizations to nonzero momentum transfer using nonuniform fields~\cite{Detmold:2004kw,Davoudi:2015cba,Bali:2015msa} have enabled studies of the axial form factor of the nucleon~\cite{Chambers:2015kuw}. Here, a new implementation 
of background fields, introduced in Ref.~\cite{Savage:2016kon}, is used to extract axial matrix elements necessary for the study of the $nn \to pp$ transition.
While $nn\to pp ee\overline{\nu}_e\overline{\nu}_e$ decay is not observed in nature because the dineutron is not bound, the nuclear matrix element is well defined within the SM and is an important subprocess in \tnubb\ decay of nuclei. It is also an important component in the \znubb-decay mode within the light Majorana-neutrino scenario. As an example of how LQCD results can provide input to many-body methods, 
the leading $\Delta I=2$ low-energy constant of pionless EFT 
(EFT($\pislash$)) is constrained from the calculated two-nucleon matrix element.
In addition to the expected Born contribution from a deuteron intermediate state, a new operator is identified
that contributes to the $\beta\beta$ decay of nuclei, but not to single-$\beta$ decays, namely the isotensor axial polarizability of the two-nucleon system. 
This contribution is determined at the unphysical quark masses used in the LQCD calculation.
If the calculations had been performed at the physical quark masses, EFT could be combined with many-body methods to 
determine the phenomenologically relevant $\beta\beta$-decay rates, 
better constraining EFT-based calculations such as those in Ref.~\cite{Menendez:2011qq}. 
Alternatively, with calculations over a range of light quark masses, an extrapolation to the physical values of the quark masses could be 
rigorously incorporated using pionful EFT.
This work demonstrates the potential  of LQCD-based approaches to address second-order electroweak properties of nuclear systems. 
With controlled systematics, future LQCD calculations of matrix elements of both short and long-distance operators 
will provide refined inputs for nuclear many-body calculations, leading to more precise predictions of both \tnubb\ and \znubb-decay rates.

\section{double-$\beta$ decay matrix elements and the isotensor axial polarizability
\label{sec:DBD-ME}}

The two-nucleon matrix elements for $2\nu\beta\beta$ decay, and $0\nu\beta\beta$ decay within a light Majorana-neutrino scenario, receive contributions from long-range second-order weak interactions. 
In both cases, the relevant nuclear matrix element is
\begin{eqnarray}
	{\cal M}_{\mu\nu}({\bm p},{\bm q},{\bm q}^\prime,E_i,E_f) & \equiv & 
	\ \frac{1}{2} \int{dt_1} {dt_2} \ 
	\langle pp;{\bm p}^\prime,E_f | T \left \{ \tilde{\mathcal{J}}^+_{\mu}({\bm q},t_1) \tilde{\mathcal{J}}^+_{\nu}({\bm q}^\prime,t_2) \right \} | nn;{\bm p},E_i \rangle,
	\label{eq:Hppnn}
\end{eqnarray}
where ${\bm p}^\prime = {\bm p}+ {\bm q} + {\bm q}^\prime$, and $E_{i,f}$ are the energies of the initial and final states. The charged weak current with three-momentum ${\bm q}$ is
\begin{eqnarray}
\tilde{\mathcal{J}}^+_\alpha({\bm q},t)=\int d^3{\bm x} \; e^{i{\bm q}\cdot{\bm x}} \ 
\overline{q}({\bm x},t) \gamma_\alpha\frac{1-\gamma_5}{2}\tau^+\ q({\bm x},t), ~ ~\text{with}~~ q=\left(
\begin{array}{ccc}
u \\
d \\
\end{array}
\right).
\end{eqnarray}
The interaction in Eq.~(\ref{eq:Hppnn}) has isospin structure $\tau^+ \otimes \tau^+$,
where $\tau^+ = \frac{1}{\sqrt{2}} \left(\tau^1 + i\; \tau^2 \right)$ and $\tau$ denotes Pauli matrices that act in isospin space, turning two down quarks into two up quarks.
In the \znubb\ transition amplitude, the contraction of 
the nuclear matrix element in Eq.~(\ref{eq:Hppnn}) with the appropriate leptonic tensor results in integration over the intermediate neutrino momentum in the nuclear matrix element. In large nuclei, the dominant contribution to such loop integrals comes from $|{\bm q}|\sim 100~\tt{MeV}$, dictated by  typical inter-nucleon distances. 
This complex process involves both the vector and axial-vector currents and is beyond the scope of the current work.
For the \tnubb\ transition amplitude, the situation is simpler as the hadronic and  leptonic matrix elements are decoupled and only phase-space integrations are 
required. Furthermore, only the Gamow-Teller (axial-vector) piece of the weak current makes a significant contribution to the decay rate 
since the long-distance contribution from the 
Fermi (vector) piece is suppressed by isospin symmetry.
Neglecting lepton-mass effects, the forward limit (${\bm q}={\bm q}^\prime=0$) of the axial-axial part of the matrix element in Eq.~(\ref{eq:Hppnn}) 
determines the \tnubb\ inverse half-life, which can be written as~\cite{Engel:2016xgb}
\begin{eqnarray}
[T^{2\nu}_{1/2}]^{-1}=
 G_{2\nu}(Q,Z) | M_{GT}^{2\nu}|^2
\ \ \ \ {\rm with}\ \ \ \
M_{GT}^{2\nu}=6\sum_{{\frak n}}\frac{\langle f |  \tilde{J}_3^+ |{\frak n}\rangle\langle {\frak n} |  \tilde{J}_3^+ |i \rangle}{E_{\frak n}-(E_{i}+E_{f})/2}.
\label{eq:MGT}
\end{eqnarray}
Here, $Q=E_i-E_f$, $Z$ is the proton number and $\tilde{J}_3^a \equiv \tilde{J}_3^a(\bm{0},t=0)  = \int d^3 \bm{x} {J}_3^a(\bm{x},t=0)$, where ${J}_3^a(x)= \overline{q}(x) \frac{\gamma_3\gamma_5}{2} \tau^a q(x)$ is the third spatial component of the $\Delta I_3=1$ zero-momentum axial current. Furthermore, ${\frak n}$ indexes a complete set of zero-momentum states and  $G_{2\nu}(Q,Z)$ is a known phase-space factor~\cite{Kotila:2012zza,Stoica:2013lka}. The factor of $6$ in $M_{GT}^{2\nu}$ is a consequence of rotational symmetry (as $M_{GT}^{2\nu}$ is written using the third spatial component of the axial currents) as well as the convention used herein for the currents. A determination of $M_{GT}^{2\nu}$ for the $nn\to pp$ transition is the focus of this work. Notably, although this transition is not observed in nature, the matrix element, and hence $M_{GT}^{2\nu}$ as defined above, are both well defined and can be determined using LQCD.

By isospin symmetry, the forward limit of the axial-axial matrix element, $M_{GT}^{2\nu}$ in Eq.~(\ref{eq:MGT}) with $| i \rangle=| nn \rangle$ and $| f \rangle= | pp \rangle$, can be related to the {\it isotensor axial polarizability}, $\beta_A^{(2)}$, of the $\si$ two-nucleon system. This polarizability is defined from $M_{GT}^{2\nu}$ by subtracting the ``Born'' term corresponding to the deuteron intermediate state,
\begin{equation}
 \frac{1}{6}M_{GT}^{2\nu} = \beta_A^{(2)} - \frac{ |\langle pp|\tilde{J}_3^+|d\rangle|^2} {\Delta}, 
\label{eq:axial-polz}
\end{equation}
where $\Delta=E_{nn}-E_d$ is the energy gap between the ground state of the isotriplet (dinucleon) and isosinglet (deuteron) channels. Note that the isotensor axial polarizability 
introduced here is unrelated to the isoscalar axial polarizability of the nucleon considered 
in the context of two-pion exchange in nuclear forces~\cite{Delorme:1976fy,Meissner:1994kn}.

In order to extract the matrix element relevant to the $\beta\beta$-decay process in the two-nucleon system, a new implementation of the LQCD background-field technique~\cite{Savage:2016kon} is employed. For the isotensor quantities considered in this work, the background field that most straightforwardly enables extraction of the desired matrix element is an isovector field proportional to $\tau^+$. For technical reasons, the calculations performed instead employ flavor-diagonal background fields. Nonetheless, the isotensor quantities of interest are still accessible in this case. This follows by noting that the particular operator in $M_{GT}^{2\nu}$ is obtained from the $++$ component of the symmetric and traceless (in isospin indices $a,b$) $I=2$ structure
\begin{equation}
	\cO^{ab}
	=
	T\left\{\frac{1}{2}
	\left(
	{J}_3^a {J}_3^b
	+
	{J}_3^b {J}_3^a
	\right)
	-
	\frac{1}{3}
	\delta^{ab} \sum_c
	{J}_3^c {J}_3^c \right\},
	\label{eq:OOab}
\end{equation}
where the coordinate dependence of the currents is not shown for brevity and $T$ denotes the time ordering of the currents. 
Matrix elements of $\cO^{ab}$ in the $I = 1$ multiplet of two-nucleon states, $\left|\si,a\right\rangle$,
can be expressed in terms of a single reduced matrix element, 
$\cM$, given by
\begin{equation}
	\langle{}^1 S_0, c | 
	\cO^{ab}
	|{}^1 S_0, d \rangle
	=
	\frac{\cM}{2}
	\left[
	\delta^{cb} \delta^{ad}
	+ 
	\delta^{ac} \delta^{bd}
	- 
	\frac{2}{3} 
	\delta^{cd} \delta^{ab}
	\right],
	\label{eq:Oab}
\end{equation}
with the normalized states, $|{}^1 S_0,a\rangle$ related to the physical states by
\begin{eqnarray}
	|nn \rangle 
	= 
	\frac{1}{\sqrt{2}} | \si, 1 \rangle - \frac{i}{\sqrt{2}} | \si, 2 \rangle,\ \ 
	| np \rangle
	= 
	| \si, 3 \rangle,\ \ 
	|pp \rangle 
	= 
	\frac{1}{\sqrt{2}} | \si, 1 \rangle + \frac{i}{\sqrt{2}} | \si, 2 \rangle
	\label{eq:states}.
\end{eqnarray}

It is clear from the isospin structure of the operator inducing the $nn \rightarrow pp$ transition that there are no self-contractions of the quark fields in the axial-current operators, 
no contractions of quark fields between the two axial-current operators, 
and no double insertions of axial-current operators on a single quark line. Since the flavor-conserving $I_3=0$ component of the operator defined in Eq.~(\ref{eq:Oab}) is most amenable to LQCD computations, it is convenient to determine the following equivalent combination of matrix elements
\begin{eqnarray}
  \langle pp | \cO^{++}|nn\rangle=
	\langle n p | 
 	\cO^{33}
 	| n p \rangle
	-
	\langle n n | 
	\cO^{33}
	| n n \rangle,
	\label{eq:6}
	\end{eqnarray}
noting that the trace subtraction, 
$-\frac{1}{3}\sum_c T\{{J}^c_3{J}^c_3\}$, 
is isoscalar and therefore cancels in the difference. It is also convenient to add to Eq.~(\ref{eq:6}) the (vanishing) difference between the matrix elements of two insertions of the isoscalar current, defined as ${S}_3=\overline{q} \frac{\gamma_3\gamma_5}{2}q$, in the $np$ and $nn$ states,
\begin{eqnarray}
	\langle pp| \cO^{++}|nn\rangle
	=
	\langle n p | 
	T\{{J}^3_3{J}^3_3\}
	| n p \rangle
	+
	\langle n p | 
	T\{{S}_3{S}_3\}
	| n p \rangle
	-
	\langle n n | 
	T\{{J}^3_3{J}^3_3\}
	| n n \rangle
	-
	\langle n n | 
	T\{{S}_3{S}_3\}
	| n n \rangle.
	\label{eq:add}
	\nonumber\\
	\end{eqnarray}
Finally, rearranging the flavor components leads to
\begin{eqnarray}
	\langle pp | \cO^{++} | nn \rangle
	&=&
	\langle np | T\{{J}^{(u)}_3{J}^{(u)}_3\} |np \rangle 
	- 
	\frac{1}{2}
	\langle n n | T\{{J}^{(u)}_3{J}^{(u)}_3\} | nn \rangle 
	- 
	\frac{1}{2}
	\langle n n | T\{{J}^{(d)}_3{J}^{(d)}_3\} |nn \rangle, 
	\label{eq:recipe}
\end{eqnarray}
where ${J}^{(f)}_3 = \overline{q}_f \gamma_3\gamma_5 q_f$
is the axial current coupled to a particular quark flavor, $f$.

\section{Axial current matrix elements from LQCD in background fields}
\label{sec:LQCD}

To determine the matrix elements relevant to $\beta\beta$ decay from LQCD, 
the fixed-order background-field approach introduced in Ref.~\cite{Savage:2016kon} is implemented.\footnote{A related method was  recently presented in Ref.~\cite{Bouchard:2016heu}.} 
Details of this method will be presented in the following section, along with the correlation functions and ratios thereof that are utilized in the analysis. Additional technical details regarding operator renormalization and finite-volume effects will be discussed at the end of the section.

\subsection{Background-field technique}
\label{sec:bf}

In the implementation of LQCD background fields in Ref.~\cite{Savage:2016kon}, hadronic correlation functions are modified directly at the level of the valence quark propagators. This is in contrast to the traditional approach where the background field modifies the action~\cite{Fucito:1982ff,Bernard:1982yu,Martinelli:1982cb,Chambers:2014qaa,Chambers:2015bka,Chambers:2015kuw}. 
In more generality than presented in Ref.~\cite{Savage:2016kon}, 
such \emph{compound propagators} in the background field can be written as
\begin{eqnarray}
S_{\{\Lambda_1,\Lambda_2,\ldots \}}(x,y) = S(x,y)
&+& \int \! dz \, S(x,z) \Lambda_1(z)  S(z,y)
\nonumber \\
&+& \int \! dz \int \! dw \, S(x,z) \Lambda_1(z) S(z,w) \Lambda_2(w) S(w,y) + \ldots
\label{eq:bfprop},
\end{eqnarray}
where both $\Lambda_i(x)$ and the quark propagator $S(x,y)$ are spacetime-dependent matrices in spinor and flavor space, while $S(x,y)$ is also a matrix in color space. 
Once the background fields $\Lambda_i(z)$ are specified, the standard sequential-source technique is used to calculate the second, third 
and all subsequent terms in Eq.~(\ref{eq:bfprop}), which are then combined to form the compound propagator. 
As implemented here, this approach is only exact for isovector fields and, even then, only for quantities that are maximally stretched in isospin space and thus do not involve operators that couple to the sea quarks. 
At the single-insertion level, this corresponds to isovector quantities such as the isovector axial charges of the proton and triton, and the axial matrix element relevant for the $pp\to d e^+\nu_e$ fusion cross section. 
With two insertions of the background field, either through the third term in Eq.~(\ref{eq:bfprop}) or from single insertions 
on two different propagators, isotensor quantities can be computed exactly. 
To compute more general quantities, the coupling of background fields to the sea quarks must be included, 
either in the generation of dynamical gauge 
configurations~\cite{Chambers:2015bka} or through reweighting methods~\cite{Freeman:2014kka}. 

In order to extract matrix elements of currents that involve zero-momentum insertion, a uniform background field is implemented. 
For the current work, a set of flavor-diagonal background axial-vector fields is used, with operator structure
\begin{eqnarray}
\Lambda^{(u)} = \lambda_u\  \gamma_3 \gamma_5 (1+\tau_3)/2
\qquad {\rm and }\qquad 
\Lambda^{(d)} = \lambda_d\  \gamma_3 \gamma_5  (1-\tau_3)/2,
\label{eq:Lambda}
\end{eqnarray}
where $\lambda_q$ are parameters specifying the strength of the background field. Zero-momentum--projected correlation functions
\begin{eqnarray}
C^{(h)}_{\lambda_u;\lambda_d}(t) 
& = & 
\sum_{\bm x}
\langle 0| \chi_h({\bm x},t) \chi^\dagger_h({\bm 0},0) |0 \rangle_{\lambda_u;\lambda_d}
\label{eq:bfcorr}
\end{eqnarray}
are formed from the compound propagators $S_{\{\Lambda^{(u)}\}}(x,y)$ and $S_{\{\Lambda^{(d)}\}}(x,y)$ that have at most a single insertion of the background field
(indicated by $\langle\ldots\rangle_{\lambda_u;\lambda_d}$). 
Here, $h$ denotes the quantum numbers of the hadronic interpolating operator, $\chi_h$. The interpolators used here are those previously utilized in the spectroscopy studies of Refs.~\cite{Beane:2012vq,Beane:2013br}. 
The correlation functions $C_{\lambda_u;\lambda_d}^{(h)}(t)$ are, by construction, polynomials of maximum degree 
$\lambda_u^{N_u}\lambda_d^{N_d}$ in the field strengths, 
where $N_{u(d)}$ is the number of up (down) quarks in the interpolating operator. 

The LQCD gauge-field configurations used in this study are the same as those used in Ref.~\cite{Savage:2016kon}. 
In particular, calculations are performed on a single ensemble of gauge-field configurations generated with a L\"uscher-Weisz gauge 
action~\cite{Luscher:1984xn} and a clover-improved fermion action~\cite{Sheikholeslami:1985ij} with $N_f = 3$ 
degenerate light-quark flavors. 
The quark masses are tuned to the physical strange quark mass, producing a pion of mass $m_\pi \approx 806~\tt{MeV}$. 
The ensemble has a spacetime volume of $L^3\times T=32^3\times48$, and a gauge coupling that corresponds to a lattice 
spacing of $a\sim 0.12~{\rm fm}$ \cite{MeinelPC}. 
For the present calculations, 437 configurations spaced by 10 Hybrid Monte Carlo trajectories are used, and 
propagators are generated from smeared sources at sixteen different locations on each configuration with both smeared (SS) 
and point (SP) sinks and at six different nonzero values of the background field-strength parameters 
$\lambda_{u,d}=\{\pm 0.05, \pm 0.1, \pm 0.2\}$, as well as at $\lambda_{u,d}=0$. 
These propagators are used to produce correlation functions for all allowed spin states of single and two-nucleon states, $h\in\{p,n,np(\siii),nn,np(\si),pp\}$. Results from different source locations are blocked on each configuration before any subsequent analysis.

\subsection{Correlation functions and matrix elements}
\label{sec:corrfns}

Both the first and second-order weak matrix elements are required for the determination of $M_{GT}^{2\nu}$. These are extracted from the response of two-point correlation functions, defined in Eq.~\eqref{eq:bfcorr}, to the background field. The first-order response to the field determines the isovector axial charge of the nucleon and the nuclear matrix element relevant for $pp\to d e^+\nu_e$, while the second-order response determines the $nn\to pp$ transition matrix element. Isolating these quantities requires a detailed analysis of the correlation functions presented in the following subsections.

In what follows, the finite temporal extent of the lattice is ignored. In principle, there are thermal contributions in which hadronic states propagate between the source and sink by going around the temporal boundary. The present analysis is confined to source-sink separations $t<T/3$, so these thermal effects are suppressed by at least $e^{-2 m_\pi T/3} \sim 10^{-7}$ relative to the dominant contributions.

\subsubsection{The Proton Axial Charge}

As the proton has two valence up quarks and one valence down quark,
the correlation function $C^{(ps)}_{\lambda_u;\lambda_d}(t)$ (where $s$ denotes the spin) 
is at most quadratic in $\lambda_u$ and linear in $\lambda_d$.
Explicitly, for a spin-up proton, and for nonzero $u$ or $d$ background axial fields, respectively,
\begin{eqnarray}
C^{(p\uparrow)}_{\lambda_u;\lambda_d=0}(t) 
&=&
\sum_{\bm x}^{}
\left(\ 
\vphantom{\sum_{\bm x}^{} }
\langle 0| \chi_{p\uparrow}({\bm x},t) \chi^\dagger_{p\uparrow}(0) |0 \rangle 
+ \lambda_u
\sum_{\bm y}\sum_{t_1=0}^t
\langle 0| \chi_{p\uparrow}({\bm x},t) J_3^{(u)} ({\bm y},t_1)  \chi^\dagger_{p\uparrow}(0) |0 \rangle \ \right)
+ d_2 \lambda_u^2, 
\nonumber 
\qquad \\
C^{(p\uparrow)}_{\lambda_u=0;\lambda_d}(t) 
& =& 
\sum_{\bm x}
\left(\ 
\vphantom{\sum_{\bm x}^{} }
\langle 0| \chi_{p\uparrow}({\bm x},t) \chi^\dagger_{p\uparrow}(0) |0 \rangle 
+ \lambda_d 
\sum_{\bm y}\sum_{t_1=0}^t
\langle 0| \chi_{p\uparrow}({\bm x},t) J_3^{(d)} ({\bm y},t_1) \chi^\dagger_{p\uparrow}(0) |0 \rangle
\ \right),
\label{eq:prot}
\end{eqnarray}
where $d_2$ is a higher-order term not needed for the present analysis. The correlation function is defined in Euclidean spacetime, and the sum over $t_1$ extends only over the temporal range between the source and the sink because of the isoscalar nature of the vacuum (exponentially small contributions that are suppressed by the mass of the lightest axial-vector meson are ignored). Given the summation over $t_1$, this procedure resembles the ``summation method'' of Ref.~\cite{Maiani:1987by}. The above expressions hold configuration-by-configuration as well as on the ensemble average. 
As a result, their polynomial structure is exact and the linear terms can be determined, given calculations of the correlation functions at at least two (three) value(s) of the field strengths $\lambda_{d(u)}$.

The coefficient of $\lambda_u$ in the first line of Eq.~(\ref{eq:prot}) is
\begin{eqnarray}
\left.
C^{(p\uparrow)}_{\lambda_u;\lambda_d=0}(t)\right|_{{\cal O}(\lambda_u)} &=& 
\sum_{{\bm x},{\bm y}}\sum_{t_1=0}^t
\langle 0| \chi_{p\uparrow}({\bm x},t) J_3^{(u)} ({\bm y},t_1)  \chi^\dagger_{p\uparrow}(0) |0 \rangle 
\nonumber \\
&=& \sum_{\frak{n},\frak{m}}\sum_{{\bm x},{\bm y}}\sum_{t_1=0}^t
\langle 0| \chi_{p\uparrow}({\bm x},t) |\frak{n} \rangle \langle \frak{n}| J_3^{(u)} ({\bm y},t_1) |\frak{m} \rangle \langle \frak{m}| \chi^\dagger_{p\uparrow}(0) |0 \rangle,
\label{eq:Olambda}
\end{eqnarray}
where ``$\big|_{{\cal O}(\lambda_q^j)}$'' denotes the piece proportional to $\lambda_q^j$ and $\frak{n}$ and $\frak{m}$ are summed over complete sets of energy eigenstates, with eigenenergies $E_{\frak{n}}$ and $E_{\frak{m}}$, respectively.\footnote{A nonrelativistic normalization of states is used throughout such that the complete set of states is $\sum_{\frak n}|{\frak n}\rangle\langle {\frak n}|=1$, and $\langle {\frak n}| {\frak m}\rangle = \delta_{{\frak m},{\frak n}}$, where $\frak{n}$ is a collective label in the case of multi-particle states.} 
Using the Hamiltonian to express the Euclidean time evolution, and performing the sum over the insertion time as an integral, which is valid up to discretization corrections, the correlation function in Eq.~(\ref{eq:Olambda}) becomes
\begin{eqnarray}
\left.
C^{(p\uparrow)}_{\lambda_u;\lambda_d=0}(t)\right|_{{\cal O}(\lambda_u)} 
&= &
\sum_{t_1=0}^t \sum_{\frak{n},\frak{m}}
z_\frak{n} z^\dagger_\frak{m} e^{-E_\frak{n}(t-t_1)} e^{-E_\frak{m}t_1} \langle \frak{n}| \tilde{J}_3^{(u)} | \frak{m} \rangle 
\nonumber \\
&= &
\sum_{\frak{n},\frak{m}}
z_\frak{n} z^\dagger_\frak{m} \frac{e^{-E_\frak{n}t}- e^{-E_\frak{m} t}}{aE_\frak{m}-aE_\frak{n}} \langle \frak{n}| \tilde{J}_3^{(u)} | \frak{m} \rangle 
 \nonumber \\
&\stackrel{t \to\infty}{\longrightarrow}&~
|z_0|^2 e^{-E_0 t} \left[ c+
t\, \langle {p\uparrow}| {\tilde{J}_3^{(u)}} | {p\uparrow} \rangle +
{\cal O}(e^{-\hat{\delta} t})
\right],
\label{eq:Olambda2}
\end{eqnarray}
where only states with zero spatial momentum and total spin equal to that of the spin-up proton contribute to the sum in the first two lines, $z_{\frak n}$ is proportional to the overlap of the interpolating operator onto a given state, i.e., $z_{\frak n}=\sqrt{V} \langle {\frak n} | \chi_{p\uparrow}(0) | 0 \rangle$, and quantities with subscript $0$ correspond to the ground state. Terms involving the time-independent constant $c$ and the leading exponential contamination  
are complicated functions of the energy gaps, excited-state overlap factors and transition matrix elements. These terms will not produce linear time dependence in the bracket in Eq.~(\ref{eq:Olambda2}) at late times. 
Similar expressions can be obtained for the spin-down state and for the response to the background field with $\lambda_u=0$ and $\lambda_d \neq 0$.
Finally, the bare isovector axial matrix element can be obtained from the late-time behavior of the difference\footnote{Note that the convention used for the axial current differs from that of Ref.~\cite{Savage:2016kon} by a factor of $\frac{1}{2}$, following the definitions after Eq.~(\ref{eq:Hppnn}).}
\begin{eqnarray}
\overline{R}_p(t) \equiv R_p(t+a)-R_p(t) ~&\stackrel{t \to \infty }{\longrightarrow}&~ \langle p | \tilde{J}_3^3 |p \rangle =\frac{g_A}{2Z_A}
\,,
\label{eq:gAeff}
\end{eqnarray}
where the ratios $R_p(t)$ are spin-weighted averages,
\begin{eqnarray}
R_p(t)
&=&\sum_{s=\{\downarrow,\uparrow\}}
\frac{\eta_s}{2}
\frac{
	\left.
	C^{(ps)}_{\lambda_u;\lambda_d=0}(t)\right|_{{\cal O}(\lambda_u)} 
	- \left. C^{(ps)}_{\lambda_u=0;\lambda_d}(t)
	\right|_{{\cal O}(\lambda_d)}
}
{ C^{(ps)}_{\lambda_u=0;\lambda_d=0}(t)},
\label{eq:Rdef}
\end{eqnarray}
with $\eta_\uparrow=-\eta_\downarrow=-1$. The factor $Z_A$ in Eq.~(\ref{eq:gAeff}) is the axial-current renormalization factor discussed in Sec.~\ref{subsec:Renorm}.

\subsubsection{$\Delta I=1$ two-nucleon axial transitions: $pp\to d e^+\nu_e$
\label{sec:3s1-1s0}}

The transition correlation functions of the $I_3=J_3=0$ two-nucleon system,\footnote{$J$ used here to represent the total angular momentum is not to be confused with the $J$ used to denote the current.} used to access the $pp$-fusion matrix 
element in Ref.~\cite{Savage:2016kon}, are at most cubic in the applied $u$ and $d$ fields. 
The forms of these correlation functions are
\begin{eqnarray}\label{eq:Csslu}
C_{\lambda_u;\lambda_d=0}^{(\siii,\si)}(t) 
&=& 
\lambda_u \sum_{t_1=0}^t \sum_{{\bm x},{\bm y}}\langle 0| \chi_{{}^3S_1}({\bm x},t) J_3^{(u)}({\bm y},t_1)\chi^\dagger_{{}{\si}}\!(0) |0 \rangle
+\
c_2 \lambda_u^2 + c_3\lambda_u^3,
\\
C_{\lambda_u=0;\lambda_d}^{(\siii,\si)}(t) 
&=& 
\lambda_d \sum_{t_1=0}^t \sum_{{\bm x},{\bm y}}\langle 0| \chi_{{}^3S_1}({\bm x},t) J_3^{(d)}({\bm y}, t_1)\chi^\dagger_{{}{\si}}\!(0) |0 \rangle
+\
b_2 \lambda_d^2 + b_3\lambda_d^3,
\end{eqnarray}
where $\chi_{{}^3S_1}$ and $\chi_{{}^1S_0}$ are interpolating operators 
for the $I_3=J_3=0$ components of the $J=1$ (isosinglet) and $J=0$ (isotriplet) two-nucleon systems, respectively. The higher-order terms in field strength, $b_i$ and $c_i$, are 
not relevant to the present calculations. 
The linear terms are isolated using polynomial fits in the applied field strengths. Labeling the $\siii$ ($\si$) eigenstates with (without) a prime, it is straightforward to show that the linear term of Eq.~\eqref{eq:Csslu} can be expressed as
\begin{eqnarray}
\left.
C^{(\siii,\si)}_{\lambda_u;\lambda_d=0}(t)\right|_{{\cal O}(\lambda_u)} 
&= &
\sum_{t_1=0}^t \sum_{\frak{n},\frak{m}}
Z_{\frak{n}^\prime} Z^\dagger_\frak{m} e^{-E_{\frak{n}^\prime}(t-t_1)} e^{-E_\frak{m}t_1} \langle \frak{n}^\prime| \tilde{J}_3^{(u)} | \frak{m} \rangle \nonumber \\
&= &
 \sum_{\frak{n},\frak{m}}
Z_{\frak{n}^\prime} Z^\dagger_\frak{m} \frac{ e^{-E_{\frak{n}^\prime} t}-e^{-E_\frak{m}t}}{aE_\frak{m}-aE_{\frak{n}^\prime}} \langle \frak{n}^\prime| \tilde{J}_3^{(u)} | \frak{m} \rangle,
\end{eqnarray}
having performed the sum over the
insertion time as an integral, which is valid up to discretization corrections. Separating ground-state contributions in the initial and/or final states leads to
\begin{eqnarray}
\left.
C^{(\siii,\si)}_{\lambda_u;\lambda_d=0}(t)\right|_{{\cal O}(\lambda_u)} 
&=&
Z_d Z_{np(\si)}^\dagger e^{-\overline{E}t}
\Bigg[
\sinh\left(\frac{\Delta t}{2}\right) \left\{ \frac{\langle d| \tilde{J}_3^{(u)} | np(\si) \rangle}{a\Delta/2} + c_-\right\}
\nonumber \\
&& \hspace{4.5cm}+ \cosh\left(\frac{\Delta t}{2}\right) c_+
 +
 {\cal O}(e^{-\tilde{\delta}\, t}) \Bigg],
\label{eq:C1s03s1}
\end{eqnarray}
where $|np(\si) \rangle$ and $|d \rangle$ refer to the ground state of the isotriplet channel and to the $J_3=0$ component of the deuteron, respectively. Here and in what follows, $Z_{\frak n'}$ and $Z_{{\frak m}}$ are the overlap factors of the source and sink interpolators onto the $\frak{n}'$ and $\frak{m}$ eigenstates of the $\siii$ and $\si$ channels, respectively, and $Z_d= Z_{0'}$, $Z_{np (\si)} = Z_0$.  The energy of the ${\frak l}'^{\rm th}$ excitation in the deuteron channel is $E_{{\frak l}'} = E_{nn}+\delta_{{\frak l}'}$,
and $E_\frak{n} = E_{nn} + \delta_\frak{n}$ is the energy of the ${\frak n}^{\rm th}$ excited state of the channel with the quantum numbers of the dinucleon (note that the energy gaps in both channels are defined relative to $E_{nn}$). Finally $\overline{E}=(E_{nn} +E_d)/2$, and $\tilde{\delta}\sim \delta_\frak{m},\delta_{\frak{n}^\prime}$ denotes a generic gap between eigenenergies of two-nucleon systems. 
Furthermore, the terms
\begin{eqnarray}
c_\pm = \sum_{\frak{m}\ne np(\si)} \frac{Z_\frak{m}^\dagger}{Z_{np(\si)}^\dagger} \frac{\langle d| \tilde{J}_3^{(u)} | \frak{m} \rangle }{a\Delta + a\delta_\frak{m}}
\pm
\sum_{\frak{n}'\ne d} \frac{Z_{\frak{n}^\prime}}{Z_d} \frac{\langle \frak{n}'| \tilde{J}_3^{(u)} | np(\si) \rangle }{a\delta_{\frak{n}^\prime}} 
\end{eqnarray}
 are $t$-independent factors involving 
energy gaps, ratios of overlap factors, and transition matrix elements between the ground and excited states. 

For arbitrary values of $\Delta$, the extraction of the desired transition matrix element from Eq. (\ref{eq:C1s03s1}) will be challenging. In the present calculation, however, the splitting is small, $a \Delta < 0.01$, which affords valuable simplifications. In the limit of exact $SU(4)$ Wigner symmetry,
$\Delta\to 0$ (the $\si$ and $\siii$ eigenstates belong to a single SU(4) multiplet in this limit) and the contribution from $c_-$ to the correlation function vanishes. Thus, after removing the leading exponential dependence by forming a ratio (see below), the ground-state transition matrix element can be extracted as the coefficient of the term linear in $t$.
Away from this limit, the extraction of the ground-state transition matrix element from the linear term is contaminated by excited states through the $c_-$ term. Although this contamination is not exponentially suppressed in time compared with the ground-state contribution, it is still expected to be small. The energy splitting $\Delta$ is small as suggested by the large-$N_c$ limit of QCD ($\Delta\sim 1/N_c^2$), while the Ademollo-Gatto theorem~\cite{Ademollo:1964sr} guarantees 
that the excited-state to ground-state matrix element is suppressed by a further power of $N_c$ relative to the ground-state to ground-state matrix element. To further reduce $SU(4)$ symmetry-breaking contamination and to assess its magnitude, one may note that in the time-reversed correlation function, i.e.,
$\left.C^{(\si,\siii)}_{\lambda_u;\lambda_d=0}(t)\right|_{{\cal O}(\lambda_u)}$, the splitting $\Delta$ is replaced with $-\Delta$, changing the sign of the contamination from the $c_-$ term. It is therefore useful to form the sum and difference
\begin{eqnarray}
\left.C^\pm_{\lambda_u;\lambda_d=0}(t)\right|_{{\cal O}(\lambda_u)} =
\frac{1}{2}\left[  
\left.C^{(\si,\siii)}_{\lambda_u;\lambda_d=0}(t)\right|_{{\cal O}(\lambda_u)}
\pm
\left.C^{(\siii,\si)}_{\lambda_u;\lambda_d=0}(t)\right|_{{\cal O}(\lambda_u)}
\right],
\end{eqnarray}
in which the residual contamination in the 
time-reversal (T) even combination of correlation functions scales as $\cO(1/N_c^4)\sim 1\%$, given the $N_c$ scalings discussed above. Additionally, the T-odd combination, $\left.C^-_{\lambda_u;\lambda_d=0}(t)\right|_{{\cal O}(\lambda_u)}$, provides a numerical estimate of the magnitude of the $\cO(1/N_c^4)$ contamination  (see Sec. \ref{sec:results}). The T-even and T-odd correlation functions for $\lambda_u=0,\lambda_d \neq 0$ can be formed similarly.

Assuming isospin symmetry, the bare $pp\to d$ matrix element can be extracted from the late-time behavior of the ratio
\begin{eqnarray}
R^\pm_{\siii,\si}(t)&=&\frac{1}{2} \frac{\left . C^\pm_{\lambda_{u};\lambda_{d}=0}(t)\right|_{{\cal O}(\lambda_u)}
	-
	\left. C^\pm_{\lambda_{u}=0;\lambda_{d}}(t)\right|_{{\cal O}(\lambda_d)} }{\sqrt{C_{0;0}^{(\siii)}(t)C_{0;0}^{(\si)}(t)}}.
\label{eq:Rpm}
\end{eqnarray}
Explicitly, $R^+$ can be used to isolate the term that is linear in $t$ in Eq.~(\ref{eq:C1s03s1}),
\begin{eqnarray}
\overline{R}^+_{\siii,\si}(t) \equiv \left[R^+_{\siii,\si}(t+a)-R^+_{\siii,\si}(t)\right] &\stackrel{t \to \infty }{\longrightarrow}& \frac{1}{Z_A}
\langle d,3| \tilde{J}_3^{+}  | pp\rangle+\mathcal{O}\left( \frac{1}{N_c^4} \right),
\label{eq:pptodeff}
\end{eqnarray}
while $\overline{R}^-_{\siii,\si}(t)$, defined analogously using $R^-_{\siii,\si}(t)$, is used to assess the size of excited-state contamination from broken Wigner symmetry.

\subsubsection{Second-order matrix elements in the dinucleon system}
\label{sec:secondorder}

The second-order axial matrix elements of the dinucleon system are the primary focus of this work. Only the $I=2$ second-order matrix elements 
can be correctly recovered from compound propagators that are computed at linear order in the axial fields, as discussed in Section \ref{sec:DBD-ME}.  
The relevant background-field correlation functions are
\begin{eqnarray}
C^{(np(\si))}_{\lambda_u;\lambda_d=0}(t) 
&=&
\sum_{\bm x}
\langle 0| \chi_{np}({\bm x},t) \chi^\dagger_{np}(0) |0 \rangle 
+ \lambda_u
\sum_{{\bm x},{\bm y}}\sum_{t_1=0}^t
\langle 0| \chi_{np}({\bm x},t) J_3^{(u)} ({\bm y},t_1)  \chi^\dagger_{np}(0) |0 \rangle 
\nonumber \\ 
&&+ \frac{\lambda_u^2}{2}
\sum_{{\bm x},{\bm y},{\bm z}}\sum_{t_1=0}^t \sum_{t_2=0}^t
\langle 0| \chi_{np}({\bm x},t) J_3^{(u)} ({\bm y},t_1) 
J_3^{(u)} ({\bm z},t_2)  \chi^\dagger_{np}(0) |0 \rangle 
+ g_3 \lambda_u^3, 
\label{eq:quad1}
\end{eqnarray}
\begin{eqnarray}
C^{(nn)}_{\lambda_u;\lambda_d=0}(t) 
&=&
\sum_{\bm x}
\langle 0| \chi_{nn}({\bm x},t) \chi^\dagger_{nn}(0) |0 \rangle 
+ \lambda_u
\sum_{{\bm x},{\bm y}}\sum_{t_1=0}^t
\langle 0| \chi_{nn}({\bm x},t) J_3^{(u)} ({\bm y},t_1)  \chi^\dagger_{nn}(0) |0 \rangle 
\nonumber \\ 
&&+ \frac{\lambda_u^2}{2}
\sum_{{\bm x},{\bm y},{\bm z}}\sum_{t_1=0}^t \sum_{t_2=0}^t
\langle 0| \chi_{nn}({\bm x},t) J_3^{(u)} ({\bm y},t_1) 
J_3^{(u)} ({\bm z},t_2)  \chi^\dagger_{nn}(0) |0 \rangle, 
\label{eq:quad2}
\end{eqnarray}
\begin{eqnarray}
C^{(nn)}_{\lambda_u=0;\lambda_d}(t) 
&=&
\sum_{\bm x}
\langle 0| \chi_{nn}({\bm x},t) \chi^\dagger_{nn}(0) |0 \rangle 
+ \lambda_d
\sum_{{\bm x},{\bm y}}\sum_{t_1=0}^t
\langle 0| \chi_{nn}({\bm x},t) J_3^{(d)} ({\bm y},t_1)  \chi^\dagger_{nn}(0) |0 \rangle 
\nonumber \\ 
&&+ \frac{\lambda_d^2}{2}
\sum_{{\bm x},{\bm y},{\bm z}}\sum_{t_1=0}^t \sum_{t_2=0}^t
\langle 0| \chi_{nn}({\bm x},t) J_3^{(d)} ({\bm y},t_1) 
J_3^{(d)} ({\bm z},t_2)  \chi^\dagger_{nn}(0) |0 \rangle 
+h_3 \lambda_d^3+h_4 \lambda_d^4. 
\label{eq:quad3}
\end{eqnarray}
The matrix elements of the two identical quark-bilinear currents involve the contractions of the currents with anti-quark (quark) pairs at the source (sink), giving rise to four possibilities, while the compound-propagator method already enforces the contractions of each quark and anti-quark pair in the source and sink through only one of the currents, reducing the possibilities to two. Thus, a factor of $\frac{1}{2}$ is required to relate the second-order terms in Eqs.~(\ref{eq:quad1})-(\ref{eq:quad3}) to the current matrix elements. The pieces of these correlation functions that are quadratic in the field strength can be determined exactly, given calculations at a sufficiently large number of values of the background axial-field strength.\footnote{Isospin symmetry equates $C^{(np(\si))}_{\lambda_u;\lambda_d=0}(t) $ and $C^{(np(\si))}_{\lambda_u=0;\lambda_d}(t)$ in the case when $\lambda_u = \lambda_d$.} The correlation function for the $nn \to pp$ transition can be formed utilizing Eq.~(\ref{eq:recipe}),
\begin{eqnarray}
C_{nn\to pp}(t)&=&2\left. C^{(np(\si))}_{\lambda_{u};\lambda_{d}=0}(t)\right|_{{\cal O}(\lambda_u^2)}
-
 \left.C^{(nn)}_{\lambda_{u};\lambda_{d}=0}(t)\right|_{{\cal O}(\lambda_u^2)}
-
\left.C^{(nn)}_{\lambda_{u}=0;\lambda_{d}}(t)\right|_{{\cal O}(\lambda_d^2)},
\label{eq:Cnnpp}
\end{eqnarray}
where the objects on the right-hand side are extracted from the compound-propagator method and the correlation function on the left-hand-side encodes the desired matrix element for the $nn \to pp$ transition. After inserting complete sets of states and using Euclidean time evolution, $C_{nn\to pp}(t)$ becomes
\begin{eqnarray}
&&C_{nn\to pp}(t) =\sum_{{\bm x},{\bm y},{\bm z}}\sum_{t_1=0}^t \sum_{t_2=0}^t
\langle 0| \chi_{pp}({\bm x},t) T\left\{J_3^{+} ({\bm y},t_1) 
J_3^{+} ({\bm z},t_2) \right\}  \chi^\dagger_{nn}(0) |0 \rangle 
\nonumber\\
&& \hspace{0.75 cm} =
{2\over a^2}\sum_{\frak{n},\frak{m}, \frak{l}'}
Z_{\frak{n}} Z^{\dagger}_{\frak{m}} e^{-E_\frak{n}t} 
\frac{\langle \frak{n}| \tilde{J}_3^{+} | \frak{l}'\rangle 
\langle \frak{l}' | \tilde{J}_3^{+} | \frak{m} \rangle }
{E_{\frak{l}'}-E_\frak{m}}
\left(\frac{e^{
		-\left(E_
		{\frak{l}'}-E_\frak{n}\right)t}-1}{E_{\frak{l}'}-E_\frak{n}}
+\frac{e^{
		\left(E_\frak{n}-E_\frak{m}\right)t}-1}{
	E_\frak{n}-E_\frak{m}}\right),
\label{eq:Cnnpp2}
\end{eqnarray}
where the summations over time have been performed as integrals (the analysis is not altered significantly if the discrete summation is used). Here, $|\frak{n} \rangle ,\, |\frak{m}\rangle$ and $|\frak{l}'\rangle$ are zero-momentum energy eigenstates 
with the quantum numbers of the
$pp$, $nn$ and deuteron systems, respectively. 
 With the assumption of isospin symmetry and in the absence of electromagnetism, which is the case for the calculations presented in this work, the $nn$ and $pp$ states are degenerate.
Eq.~\eqref{eq:Cnnpp2} resembles a second-order weak correlation function calculated in the kaon system in Ref.~\cite{Christ:2012se}.

In order to make the matrix element between ground-state dinucleons explicit, the sums over states in this correlation function are partially expanded, giving
\begin{eqnarray}
&&a^2C_{nn\to pp}(t) = 2 Z_{pp}Z_{nn}^{\dagger} e^{-E_{nn}t}\ 
\Bigg\{\ 
\left[ 
{e^{\Delta t}-1\over \Delta^2}\ -\ {t\over \Delta}
\right] 
\langle pp| \tilde{J}_3^{+} |d\rangle \langle d | \tilde{J}_3^{+} | nn \rangle
\nonumber\\
&&\hspace{0.75 cm} +\left.
\sum_{{\frak l}' \ne d}
 \left[ {t\over\delta_{{\frak l}'}} - {1\over\delta_{{\frak l}'}^2}\ \right]
\langle pp| \tilde{J}_3^{+} |{\frak l}'\rangle \langle {\frak l}' | \tilde{J}_3^{+} | nn \rangle
\right.\nonumber\\
&&\hspace{0.75 cm} +
\left.
\sum_{{\frak n}\ne nn,pp} 
\left[ 
{e^{\Delta t} \over \Delta (\Delta+\delta_{\frak n} )}
- {1\over\Delta \delta_{\frak n}}
\right]
\left(\frac{Z_{\frak n}}{Z_{pp}}
\langle {\frak n} | \tilde{J}_3^{+} |d\rangle \langle d | \tilde{J}_3^{+} | nn \rangle
+
\frac{Z_{\frak n}^\dagger}{Z_{nn}^\dagger}\langle pp | \tilde{J}_3^{+} |d\rangle \langle d | \tilde{J}_3^{+} | {\frak n} \rangle
\right)
\right.\nonumber\\
&&\hspace{0.75 cm} +
\left.
\sum_{{\frak n}\ne nn,pp} 
\sum_{{\frak l}'\ne d} 
{1\over \delta_{{\frak l}'} \delta_{\frak n}}
\left(
\frac{Z_{\frak n}}{Z_{pp}}\langle {\frak n} | \tilde{J}_3^{+} |{\frak l}'\rangle \langle {\frak l}' | \tilde{J}_3^{+} | nn \rangle
+
\frac{Z_{\frak n}^\dagger}{Z_{nn}^\dagger}\langle pp | \tilde{J}_3^{+} |{\frak l}'\rangle \langle {\frak l}' | \tilde{J}_3^{+} | {\frak n} \rangle
\right)
\right.\nonumber\\
&&\hspace{0.75 cm} +
\sum_{{\frak n, \frak m}\ne nn,pp} 
 {e^{\Delta t} \over (\Delta+\delta_{\frak n})(\Delta+\delta_{\frak m}) }
 \frac{Z_{\frak n}}{Z_{pp}} \frac{Z_{\frak m}^\dagger}{Z_{nn}^\dagger}
 \langle {\frak n} | \tilde{J}_3^{+} |d\rangle \langle d | \tilde{J}_3^{+} | {\frak m} \rangle
+\cO(e^{-\delta t},e^{-\delta^\prime t})
\Bigg \}.
\label{eq:cnnppexplicit}
\end{eqnarray}
The energies and overlap factors are defined as in the previous section, see the discussion after Eq.~(\ref{eq:C1s03s1}). To arrive at Eq.~(\ref{eq:cnnppexplicit}), the deuteron-dineutron energy splitting is assumed to be modest compared with the inverse of the time separation between the source and the sink used to extract the matrix elements, while the energy splittings between ground and exited states in both channels are assumed to be large, so that
$e^{-\delta_{{\frak l}'} t}\rightarrow 0$ and $e^{-\delta_{\frak n} t}\rightarrow 0$.
If this is not the situation, the correlation functions with background-field insertions on all timeslices cannot be used to unambiguously extract the terms relevant for this analysis.\footnote{Inserting the background field on a range of timeslices  separated from the source and sink can address this issue \cite{Christ:2012se}, provided the separation is sufficiently large.} In the numerical calculations discussed below, the requisite hierarchy is found to be satisfied.
As the deuteron is lower in energy than the dinucleon external states, and hence gives rise to a growing exponential 
contribution (after the overall exponential $e^{-E_{nn}t}$ is factored out of Eq.~(\ref{eq:cnnppexplicit})),
this contribution has been singled out in the summation over states in Eq.~\eqref{eq:cnnppexplicit}.
The deuteron contribution is close to quadratic in $t$ (it would be exactly quadratic if $\Delta=0$), 
and 
the coefficient of this term is known from the first-order axial response in Eq.~(\ref{eq:pptodeff}). 
Ground-state overlap factors and the overall exponential time dependence can be removed by forming the ratio
\begin{eqnarray}
\mathcal{R}_{nn\to pp}(t)&=&\frac{C_{nn\to pp}(t)}{2 C^{(nn)}_{0;0}(t)},
\label{eq:Rnnpp}
\end{eqnarray}
which will be investigated in Sec. \ref{sec:results}.
Using Eq.~(\ref{eq:cnnppexplicit}), it is easy to show that this ratio has the form
\begin{eqnarray}
a^2\mathcal{R}_{nn\to pp}(t)
&=&
\left[ -t+
{e^{\Delta t}-1\over\Delta}
\right] 
\frac{\langle pp| \tilde{J}_3^{+} |d\rangle \langle d | \tilde{J}_3^{+} | nn \rangle}{\Delta}
\ +\ 
t \ \sum_{{\frak l}' \ne d}
 {
\langle pp| \tilde{J}_3^{+} |{\frak l}'\rangle \langle {\frak l}' | \tilde{J}_3^{+} | nn \rangle
\over \delta_{{\frak l}'}}
\nonumber\\
& & \hspace{5.75 cm} \ +\ 
C \ +\ D \ e^{\Delta t} +\cO(e^{-\delta t},e^{-\delta^\prime t}),
\label{eq:funcform}
\end{eqnarray}
where the first term is the long-distance contribution to the matrix element from the deuteron intermediate state and the second term is the short-distance contribution arising from all excited intermediate states coupling to the axial current, i.e., the isotensor axial polarizability as defined in Eq.~(\ref{eq:axial-polz}). The coefficients $C$ and $D$ are complicated terms involving ground-state and excited-state overlap factors and matrix elements, as can be read from Eq.~(\ref{eq:cnnppexplicit}), but have no time dependence. The critical aspect of Eq.~(\ref{eq:funcform}) is that both the short-distance and the 
long-distance contributions can be isolated from the excited external-state contributions through their distinct dependence on time. This form will be used to analyze the numerical correlation functions in Section \ref{sec:results}.

\subsection{Finite-volume effects}
\label{subsec:FV}

The initial and final states in the $nn\to pp$ transition are deeply bound degenerate states at the $SU(3)$ flavor-symmetric set of quark masses used in this work, which considerably simplifies the analysis. 
In addition, the dominant intermediate state that propagates between the two currents is the deuteron, which is close in 
energy to the $nn$ and $pp$ states, with no other intermediate states able to go on shell at the kinematic threshold. 
As the deuteron is also a compact bound state in this calculation, there is no complication with regard to finite-volume two-particle states and only exponentially small volume effects are anticipated. 
A similar problem has been studied in detail in the case of long-distance contributions to the $K_L$--$K_S$ 
mass difference~\cite{Christ:2015pwa}. There, however, a tower of intermediate two-pion states with energies lower than the initial-state kaon must be dealt with explicitly, introducing power-law corrections to the relation between the infinite-volume and finite-volume matrix elements (see also the related discussions of the rare weak processes $K\to\pi\nu\overline{\nu}$~\cite{Christ:2016eae,Bai:2017fkh} 
and $K\to\pi\ell^+\ell^-$~\cite{Christ:2015aha,Christ:2016mmq}). Such calculations will become increasingly difficult as the large volume limit is approached. As the present calculations of two-nucleon matrix elements are extended to lighter quark masses approaching their physical values, the initial and final states will become unbound, further complicating the extraction of infinite-volume Minkowski-space matrix elements from the Euclidean-space correlation functions. 
Such an extraction will require the use of generalized Lellouch-L\"uscher relations~\cite{Lellouch:2000pv,Briceno:2012yi,Agadjanov:2014kha,Briceno:2015tza,Christ:2015pwa}.
Eventually, the inclusion of electromagnetism will shift the single and two-nucleon spectra and will introduce Coulomb repulsion between the final-state protons, requiring extensions of the formalism developed in Refs.~\cite{Beane:2014qha,Borsanyi:2014jba,Davoudi:2014qua} to extract the physical matrix elements.

\subsection{Operator renormalization
\label{subsec:Renorm}}

The axial-current renormalization factor 
{$Z_A=0.867(43)$} was determined in Ref.~\cite{Savage:2016kon} from computations of the vector-current renormalization factor in the proton by noting that $Z_A=Z_V + {\cal O}(a)$ and assigning a $5\%$ systematic 
uncertainty associated with lattice-spacing artifacts (statistical uncertainties are negligible). This determination is used in the current work when needed. A more sophisticated determination that removes the leading lattice-spacing artifacts leads to 
$Z_A=0.8623(01)(71)$~\cite{GreenLatt2016,Yoon:2016jzj} on an ensemble with the same form of action and gauge coupling as used in this work but
at a pion mass of $m_\pi\sim 317~\tt{MeV}$.

\begin{figure}[!t]
\includegraphics[width=0.625\textwidth]{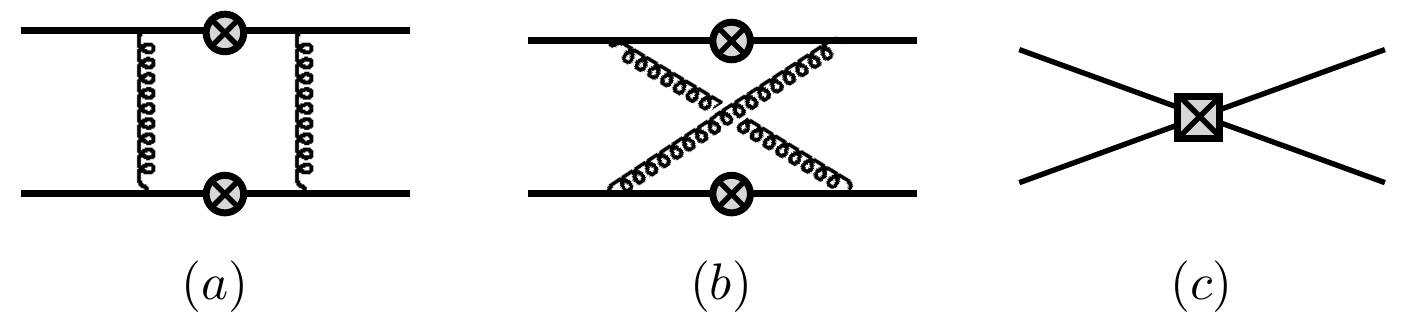}
\caption{\label{fig:Op-norm} 
The double axial-current insertion is renormalized by contributions from four-quark operators through the ultraviolet behavior of gauge interactions.
Diagrams ($a$) and ($b$) show the possible two-gluon exchanges between axial-current insertions (gray crossed circles) on the quark propagators. 
The ultraviolet behavior of these diagrams generates four-quark operators in the Szymanzik action~\cite{Symanzik:1983dc,Symanzik:1983gh}, depicted as a gray crossed square in diagram ($c$), with coefficients that scale as $\mathcal{O} (\alpha_s^2 a^2)$ near the continuum limit.}
\end{figure}

While double insertions of the axial current generally renormalize straightforwardly as products of two axial-current insertions, 
additional care is required for contributions where both insertions become localized around the same spacetime point, 
which necessarily occurs in this background-field approach 
(for zero-momentum background fields, these contributions are naively suppressed by the spacetime volume). 
Because of interactions, 
such contributions are no longer proportional to the product of two axial currents. 
In particular, four-quark operators are radiatively generated in the context of Symanzik's effective action~\cite{Symanzik:1983dc}.
Such short-distance contributions are shown in 
Fig.~\ref{fig:Op-norm}, 
and arise from the ultraviolet behavior of diagrams involving the exchange of at least two gluons between the axial-current insertions. 
In the case of the isotensor operator, there are thus lattice-spacing artifacts arising from  four-quark operators such as
$\mathcal{Q}^{ab} = (\overline{q} \frac{\tau^a}{2} \gamma_3 \gamma_5 q) (\overline{q} \frac{\tau^b}{2} \gamma_3 \gamma_5 q)$, 
where the isospin indices $a$ and $b$ require symmetrization and trace subtraction. 
The mixing coefficients governing the renormalization of the full set of four-quark operators scale with $\alpha_s^2 a^2$, and are hence expected to 
yield sub-percent contributions that can be neglected in this analysis. 
As a result, the isotensor axial polarizability can be renormalized by $Z_A^2$. These renormalization factors cancel, moreover, in the ratio of the polarizability to the square of the single-nucleon axial coupling.

%

\section{Analysis of the Lattice Calculations}
\label{sec:results}
 
In this section, details of the analysis of the numerical lattice calculations are presented, along with results for the matrix elements discussed in Sec.~\ref{sec:corrfns}. For each of the correlation functions discussed in Sec.~\ref{sec:corrfns}, the computed values from all 16 source locations on a given configuration are first averaged to produce one value for each of the 437 configurations. 
These averaged values are then resampled using a bootstrap procedure, with the variation over the bootstrap ensembles propagated to define the statistical uncertainty of all derived quantities. Systematic uncertainties are addressed by considering the choice of temporal fit ranges, higher-order terms (where appropriate), and additionally from the comparison of multiple independent analyses in which specific details of the fit procedures are different. In what follows, figures from a single analysis are presented, but the final numerical values include this additional uncertainty.

\begin{figure}[!t]
\includegraphics[width=0.95\textwidth]{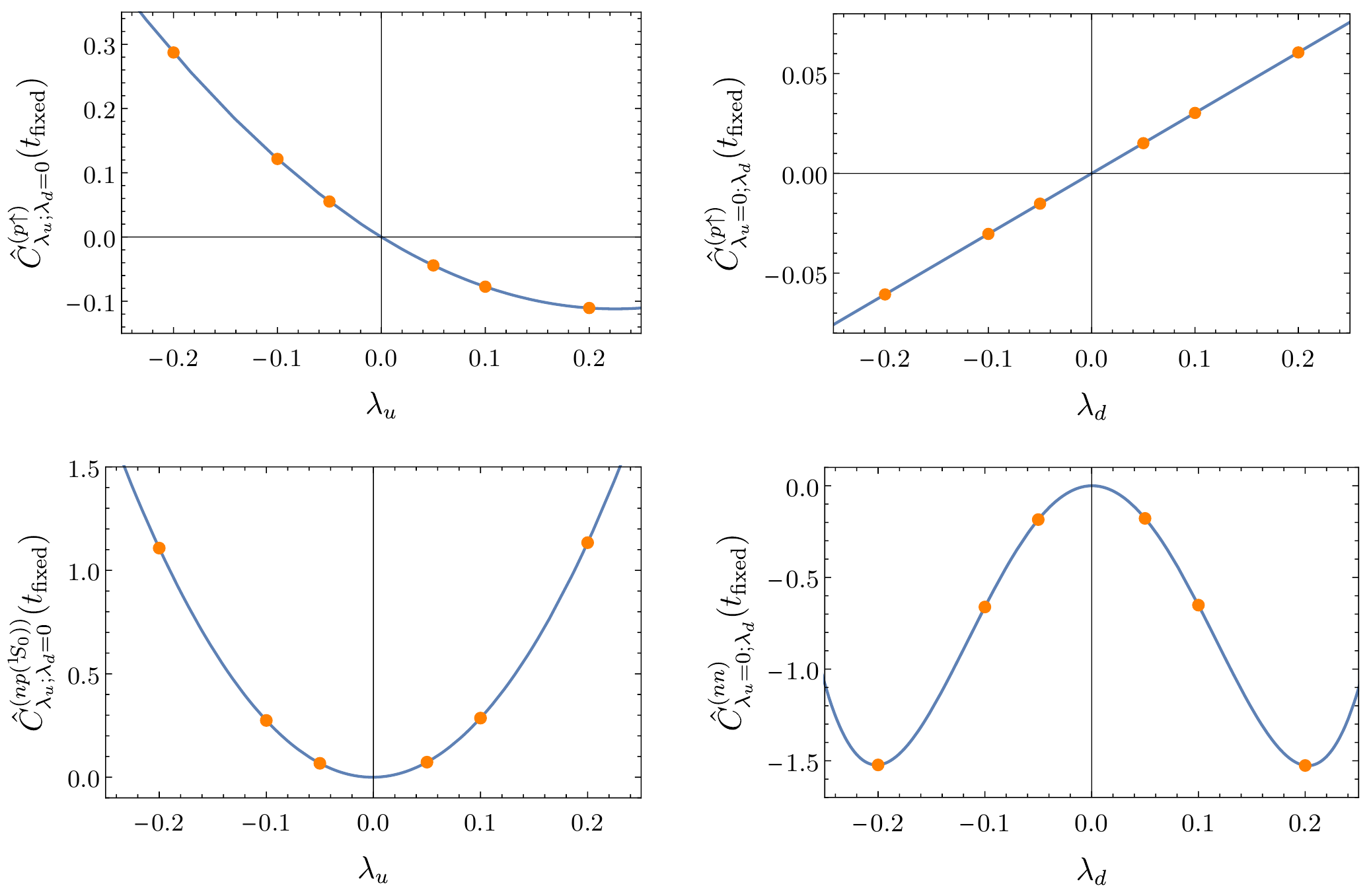}
\caption{
The field-strength dependence of sample correlation functions constructed from compound propagators on a given configuration at a given time (each configuration and timeslice shows similar polynomial behavior). 
The quantities shown are correlation functions with the zero-field limit subtracted: $\hat{C}^{(h)}_{\lambda_u;\lambda_d}(t)=C^{(h)}_{\lambda_u;\lambda_d}(t)-C^{(h)}_{\lambda_u=0;\lambda_d=0}(t)$.
The solid curves show the polynomials used to extract the requisite linear and quadratic responses. The points denote the results of numerical calculations at six values of the field strength.
}
\label{fig:field_response}
\end{figure}

To determine the matrix elements of interest from the hadronic correlation functions, these functions must be separated into linear, quadratic and higher powers of insertions of the up-quark and down-quark axial-current operators, as described in Sec.~\ref{sec:corrfns}. 
In Fig.~\ref{fig:field_response}, the field-strength dependence of representative correlation functions is shown at a given timeslice and on a particular gauge-field configuration, along with the polynomial forms that enable the extraction of the linear and quadratic responses. 
As discussed previously, with the number of field strengths being greater than or equal to the number of terms in the polynomial, the fit is a direct solution. 
\begin{figure}[!t]
    \includegraphics[width=1.0\columnwidth]{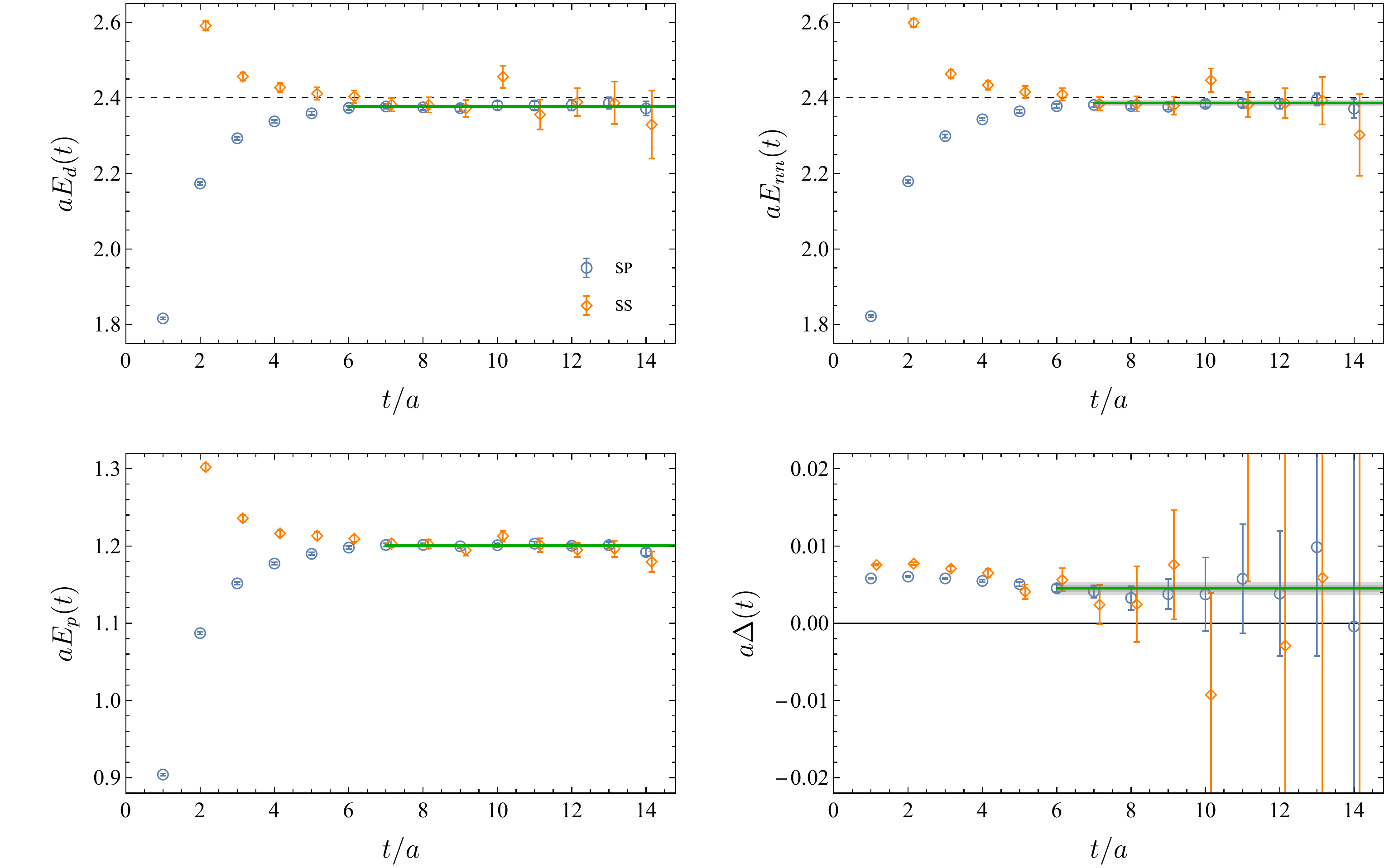}
\label{fig:EMPs}
	\caption{Effective-mass plots for the deuteron (upper-left panel), dineutron (upper-right panel), nucleon (lower-left panel), and the quantity $\Delta=E_{nn}-E_d$ (lower-right panel). Blue circles and orange diamonds denote results determined using SP and SS correlation functions, respectively. The dashed lines in the upper-panel plots correspond to twice the mass of the nucleon. In all figures, the horizontal bands show constant fits to the late-time behavior of the SP quantities. The SS points are slightly offset in $t$ for clarity.}
	\label{fig:effmass}
\end{figure}
With the required linear and quadratic field-strength dependences of the correlation functions determined, the remaining task is to isolate the matrix elements of interest through the time dependence of the combinations of correlation functions derived in Sec.~\ref{sec:corrfns}. As the first-order responses have been presented in Ref.~\cite{Savage:2016kon}, the primary focus of this work is the second-order axial matrix element describing the $nn\rightarrow pp$ transition, as discussed in Sec.~\ref{sec:secondorder}.
For this matrix element, the challenge is to isolate both its long-distance and short-distance components. Since the long-distance contribution can be determined 
from numerical calculations of the matrix element associated 
with a single insertion of the axial current, it can be removed from
$\mathcal{R}_{nn\to pp}(t)$ in Eq.~(\ref{eq:funcform}) to leave 
\begin{eqnarray}
\hat{\mathcal{R}}_{nn\to pp}(t)
& = & 
\mathcal{R}_{nn\to pp}(t) - \frac{|\langle pp| \tilde{J}_3^{+} |d\rangle|^2 }{a\Delta} 
\left[- \frac{t}{a} + \frac{e^{\Delta t}-1}{a\Delta}\right]
\nonumber\\
& = & \frac{t}{a} ~ \sum_{{\frak l}' \ne d}
 {
\langle pp| \tilde{J}_3^{+} |{\frak l}'\rangle \langle {\frak l}' | \tilde{J}_3^{+} | nn \rangle
\over a\delta_{{\frak l}'}}
+ 
c + d\ e^{\Delta t}.
\label{eq:rnnppsub}
\end{eqnarray}
This subtraction is most effectively done in a correlated manner, requiring determinations of the energy splitting and the $pp \to d$ matrix element.

Plots of the effective-mass functions of the nucleon, deuteron, dineutron, and of the difference $\Delta=E_{nn}-E_d$ are shown in Fig.~\ref{fig:EMPs}, along with fits to the late-time behavior of the appropriate ratios of the correlation functions. This figure shows that the deuteron and dinucleon zero-field correlation functions are saturated by their ground-state contributions by timeslice $6$. Consequently, in the ratio $\mathcal{R}_{nn\to pp}(t)$ and derived quantities, fits can only be performed over timeslices equal to or larger than this threshold, even though the ratios may appear to plateau earlier.

\begin{figure}[!t]
	\centering
	\includegraphics[width=0.985\columnwidth]{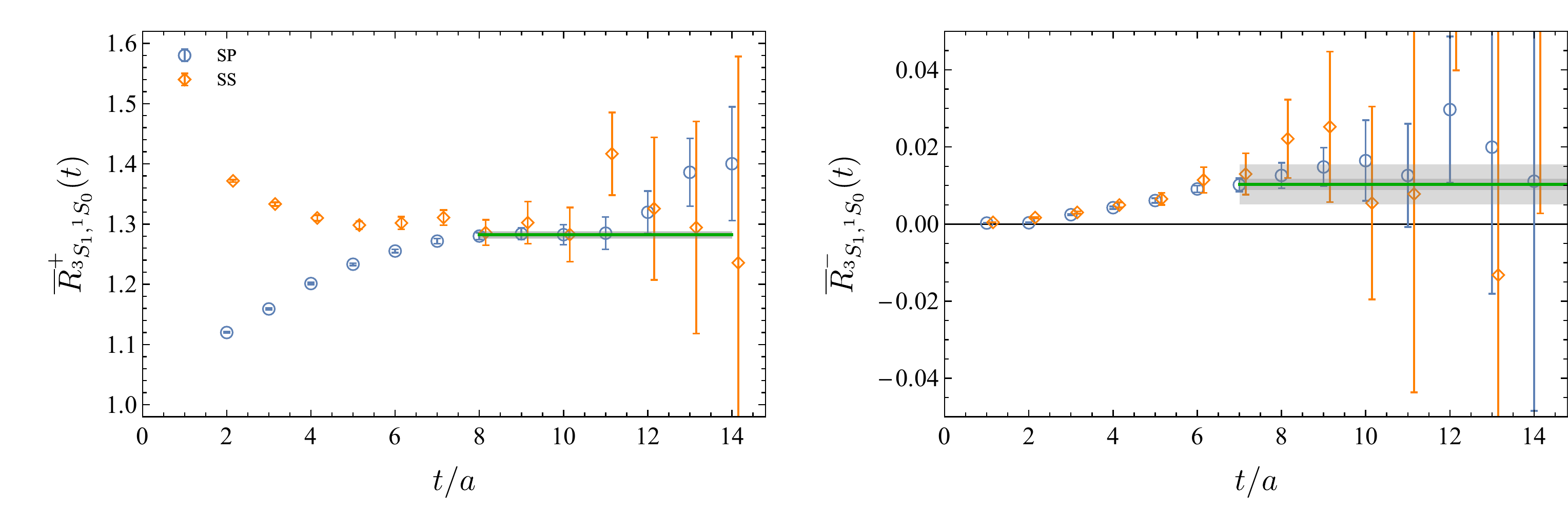}
\caption{The left panel shows the quantity $\overline{R}^+_{\siii,\si}(t)$ used to extract the bare $pp\to d$ transition matrix element. The right panel shows $\overline{R}^+_{\siii,\si}(t)$, used to estimate the magnitude of excited-state contamination in the extraction of the $pp\to d$ bare matrix element (see Sec.~\ref{sec:3s1-1s0}). Blue circles and orange diamonds denote results obtained using SP and SS correlation functions, respectively. The horizontal bands show constant fits to the late-time behavior of the SP quantities. The SS points are slightly offset in $t$ for clarity.}
	\label{fig:pptod}
\end{figure}

The quantity $\overline{R}^+_{\siii,\si}(t)$,  defined in Eq.~(\ref{eq:pptodeff}), is shown in the left panel of Fig.~\ref{fig:pptod}, along with a fit to this quantity at late times which is used to determine the value of the $pp\rightarrow d$ axial transition matrix element. In addition, the quantity $\overline{R}^-_{\siii,\si}(t)$, used to estimate the effects of excited states contaminating the extraction of the $pp\to d$ transition matrix element, is shown in the right panel of Fig.~\ref{fig:pptod}. The late-time behavior of this quantity saturates to a very small value indicating that the $N_c$ scaling is borne out (recall from Sec.~\ref{sec:3s1-1s0} that this quantity vanishes as $1/N_c^4$ based on a large-$N_c$ analysis). With this supporting evidence, it is reasonable to conclude that the contaminating term $c_-$ in Eq.~(\ref{eq:C1s03s1}) is $\cO(1/N_c^4) \sim \cO(1\%)$ of the dominant term. To account for this systematic effect, an additional Wigner symmetry-breaking uncertainty of this size is added to the value of the bare $\langle d|\tilde{J}_3^+|pp \rangle$ matrix element extracted from the late-time asymptote of $\overline{R}^+_{\siii,\si}(t)$.

\begin{figure}[!t]
\centering
\includegraphics[width=1\columnwidth]{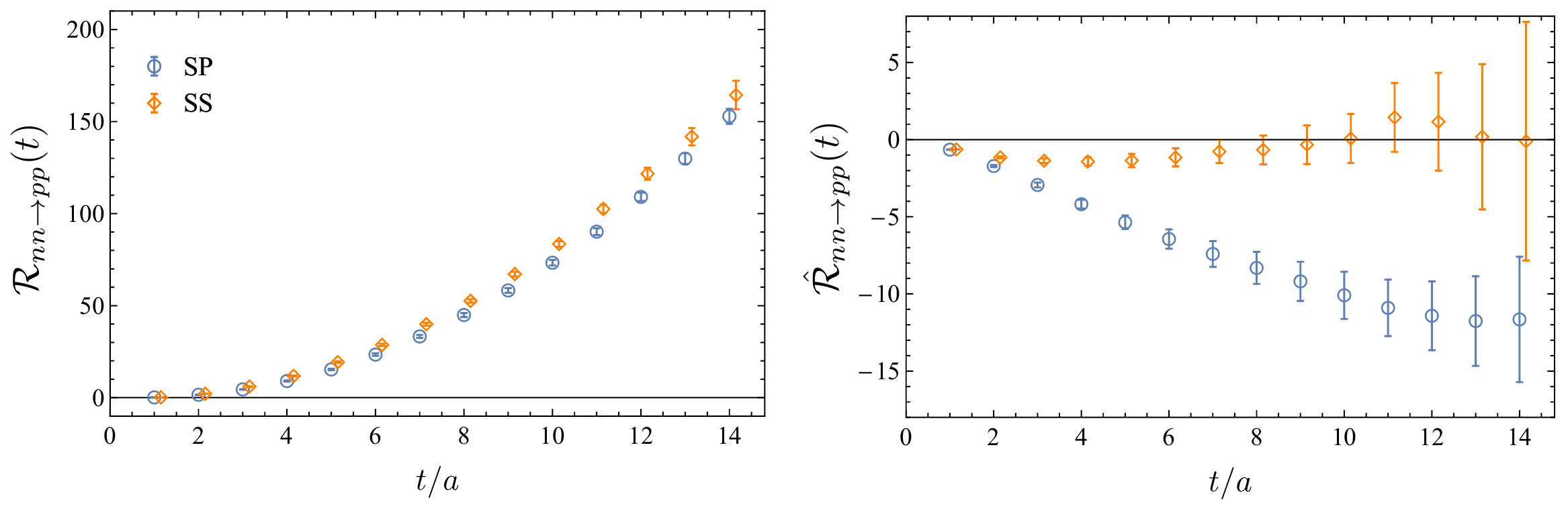}
\caption{The ratio $\mathcal{R}_{nn\to pp}(t)$ (left panel) and the subtracted ratio $\hat{\mathcal{R}}_{nn\to pp}(t)$ (right panel) that are constructed from the SP and SS
correlation functions as prescribed in Eqs.~(\ref{eq:Rnnpp})~and~(\ref{eq:rnnppsub}). Blue circles and orange diamonds denote results determined using SP and SS correlation functions, respectively. The SS points are slightly offset in $t$ for clarity.
}
\label{fig:nnppME}
\end{figure}

Fits to both the mass difference, $\Delta$, and to the bare $pp \to d$ matrix element on each bootstrap ensemble allow for the deuteron-pole term to be determined and subtracted in a correlated manner (in all cases, the statistically cleaner SP results are used for this subtraction in the results shown below). The results obtained for $\mathcal{R}_{nn\to pp}(t)$ and $\hat{\mathcal{R}}_{nn\to pp}(t)$ are shown in Fig.~\ref{fig:nnppME} for both the SS and SP source-sink combinations, demonstrating that the subtracted long-distance contribution is the dominant piece of the correlation function and provides the largest contribution to its curvature (note the different scales on the plots). If the deuteron-pole term is not directly subtracted, fits to the full time dependence of Eq.~(\ref{eq:funcform}) can be performed. Such fits result 
 in a value for the $pp\to d$ matrix element that is consistent with that obtained from the linear response of the corresponding $pp \to d$ correlation function, albeit with larger uncertainties. 
 The SP subtracted ratio is almost completely linear, indicating that, for this source--sink combination, the $C$ term in Eq.~(\ref{eq:funcform}) is very small. In contrast, the SS subtracted ratio exhibits significant nonlinearity, indicating that the $C$ term is larger in this case. This behavior is in accordance with expectations; the SP sink has a highly suppressed overlap onto the $nn$ scattering states that dominate the excitations that contribute to the $C$ term.
  
\begin{figure}[!t]
\centering
\includegraphics[width=1\columnwidth]{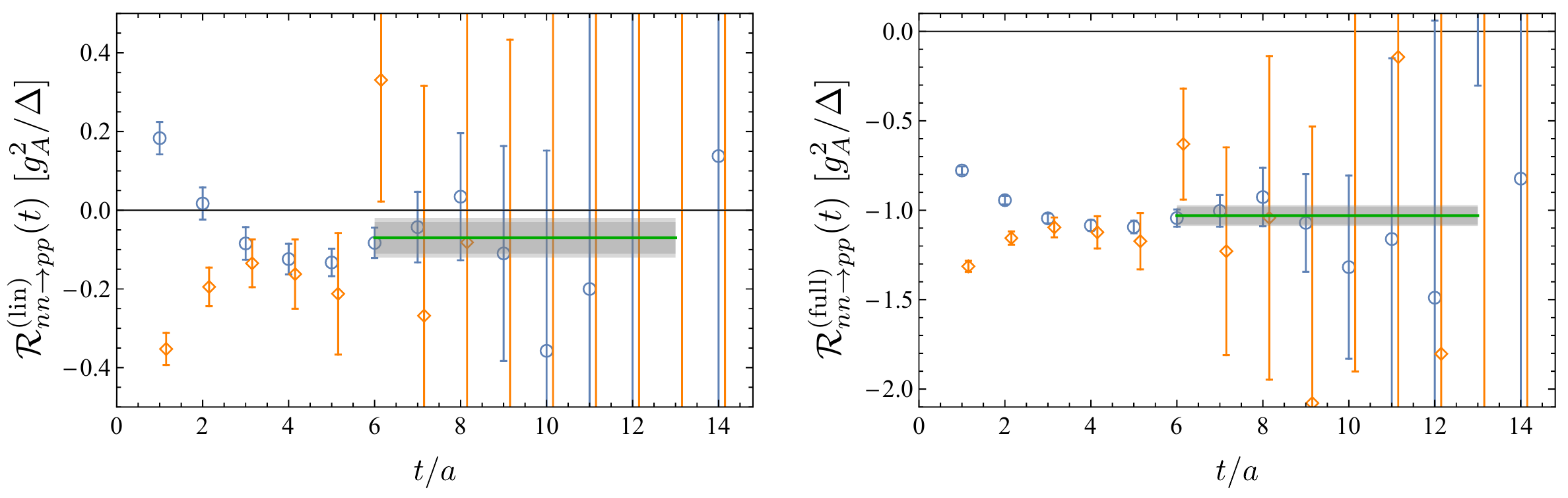}
\caption{The left panel shows $\mathcal{R}^{\text{(lin)}}_{nn\to pp}(t)$ (normalized by $g_A^2/\Delta$), corresponding to the bare short-distance contribution to the $nn \to pp$ matrix element at late times,  Eq.~(\ref{eq:Rlin}). The right panel shows $R^{\text{(full)}}_{nn\to pp}(t)$ (normalized by $g_A^2/\Delta$), which sums the long-distance and short-distance contributions to the matrix element, Eq.~(\ref{eq:Rfull}). 
In both panels, the orange diamonds and blue circles correspond to the SS and SP results, respectively. The horizontal bands denote constant fits to the SP results at late times, which are used to extract the final values of the matrix elements. The SS points are slightly offset in $t$ for clarity.
}
\label{fig:MEextract}
\end{figure}

In order to separate the desired short-distance contribution to $\hat{\mathcal{R}}_{nn\to pp}(t)$
from the effects of excited external states that couple to the source and sink, the linear $t$ dependence of $\hat{\mathcal{R}}_{nn\to pp}(t)$ must be distinguished from exponential contributions of the form $e^{\Delta t}$. This separation can be accomplished straightforwardly by forming the following combination of $\hat{\mathcal{R}}_{nn\to pp}$ at three neighboring timeslices:
\begin{eqnarray}
\mathcal{R}^{\text{(lin)}}_{nn\to pp}(t) =
\frac{(e^{a\Delta} +1) \hat{\mathcal{R}}_{nn\to pp}(t+a) - \hat{\mathcal{R}}_{nn\to pp}(t+2a) - e^{a\Delta} \hat{\mathcal{R}}_{nn\to pp}(t)}{e^{a\Delta} -1}
~\stackrel{t \to \infty }{\longrightarrow}~ \frac{1}{aZ_A^2}\frac{\beta_A^{(2)}}{6}.
\label{eq:Rlin}
\end{eqnarray}
As denoted, at large time separations, $\mathcal{R}^{\text{(lin)}}_{nn\to pp}(t)$ asymptotes to the bare isotensor axial polarizability, as defined in Eq.~(\ref{eq:axial-polz}).
This term can now be combined with the deuteron-pole contribution in a correlated manner to form 
\begin{eqnarray}
\mathcal{R}^{\text{(full)}}_{nn\to pp}(t) = \mathcal{R}^{\text{(lin)}}_{nn\to pp}(t) - \frac{|\langle pp| \tilde{J}_3^{+} |d\rangle|^2 }{a\Delta}
~\stackrel{t \to \infty }{\longrightarrow}~ \frac{1}{aZ_A^2}\frac{M^{2\nu}_{GT}}{6}, 
\label{eq:Rfull}
\end{eqnarray}
which asymptotes to the bare Gamow-Teller matrix element.
The results for both $\mathcal{R}^{\text{(lin)}}_{nn\to pp}(t)$ and $\mathcal{R}^{\text{(full)}}_{nn\to pp}(t)$ are shown in Fig.~\ref{fig:MEextract}, along with fits to the asymptotic behavior of the  SP correlation functions. Constant behavior is observed at late times, with consistent results from the SS and (significantly more precise) SP combinations, an indication that the assumptions made in deriving Eq.~(\ref{eq:cnnppexplicit}) are valid. An estimate of the finite-volume excited-state spectra of the isosinglet and isotriplet two-nucleon systems based on the phase shifts extracted in Ref.~\cite{Beane:2013br} further validates the assumed hierarchy of the ground and excited-states gaps, and numerically shows that $\delta\sim 8\Delta$.

The fits to the SP effective matrix elements shown in Fig.~\ref{fig:MEextract} yield the following values of the long-distance, short-distance and total matrix elements for 
$nn\rightarrow pp$ transition resulting from two insertions of the axial current:
\begin{eqnarray}
{\Delta\over g_A^2} \frac{|\langle pp| \tilde{J}_3^{+} |d\rangle|^2 }{\Delta}
& = & 
1.00(3)(1),
\\
{\Delta\over g_A^2}\sum_{{\frak l}' \ne d}
 {
\langle pp| \tilde{J}_3^{+} |{\frak l}'\rangle \langle {\frak l}' | \tilde{J}_3^{+} | nn \rangle
\over \delta_{{\frak l}'}}
& = & 
-0.04(4)(2),
\\
{\Delta\over g_A^2}\frac{M_{GT}^{2\nu}}{6} &=&
-1.04(4)(4),
\end{eqnarray}
where in order to suppress the $\cO(a)$ lattice-spacing artifacts from the axial currents, the quantities are normalized by $g_A^2/\Delta$ in a correlated manner to produce combinations that are independent of the axial-current renormalization constant, $Z_A$.
In each of these expressions, the first uncertainties arise from statistical sampling, systematic effects from fitting choices, and deviations from Wigner symmetry as described in Sec.~\ref{sec:3s1-1s0}. The second uncertainties encompass differences between analysis methods. Clearly, the short-distance contribution is suppressed relative to the deuteron-pole contribution but it is non-negligible.
There are additional systematic uncertainties that are not included in the above uncertainty estimations, including finite-volume effects, lattice-spacing artifacts, and electromagnetic and quark-mass effects. At present, it is difficult to quantify such uncertainties, although they are not expected to qualitatively alter the results of this exploratory calculation. In the future, it will be important to investigate such effects by improving upon the  calculations presented here, as discussed further in Sec.~\ref{sec:sum}.
%

\section{Second-order weak processes in pionless EFT}
\label{sec:pionless}

In this section, the results of the LQCD calculations are matched to EFT($\pislash$) 
and explicitly used to determine the coefficient of a short-distance, two-nucleon, second-order axial-current operator in the dibaryon formalism. 
In principle, with this contribution constrained, EFT($\pislash$) can be used to calculate $\beta\beta$-decay rates of light nuclei at this pion mass.
EFT($\pislash$)~\cite{Kaplan:1996nv,Kaplan:1998tg, Kaplan:1998we, vanKolck:1998bw, Chen:1999tn, Beane:2000fi} 
is a natural approach to use at this heavy quark mass
as the momenta involved in  $2\nu\beta\beta$ decays are small compared with the start of the two-nucleon $t$-channel cut when 
isospin breaking and electromagnetism are included (in this isospin-symmetric numerical work, the transition is below threshold for massive leptons).
At lighter quark masses, including the physical point, pionful EFTs will likely be required~\cite{Davoudi:2017}.

\subsection{Review of pionless EFT in the dibaryon approach}
\label{pislashEFT}

At momenta well below the pion mass, $|p| \ll m_{\pi}$, the strong interactions of two-nucleon systems, 
as well as their interactions with background fields, can be systematically studied in the framework of 
EFT($\pislash$)~\cite{Kaplan:1998tg, Kaplan:1998we, Chen:1999tn, Beane:2000fi}. 
As $s$-wave interactions in the two-nucleon sector are strong, generating anomalously large two-nucleon scattering lengths, they must be included to all orders.
However, interactions in higher partial waves can be included perturbatively. 
In the dibaryon formulation of  EFT($\pislash$)~\cite{Beane:2000fi, Phillips:1999hh}, this resummation is accomplished by dressing the $s$-wave dibaryon propagators, by including $s$-channel rescattering to all orders. 
In terms of the nucleon field, $N$, and the isosinglet ($\siii$) and isotriplet ($\si$) dibaryon fields, $t_i$ and $s_a$, the Lagrangian in the absence of background fields can be written as 
\begin{eqnarray}
{\cal L}^{(0)} & = & N^\dagger\left[ i\partial_0 + {\nabla^2\over 2 M} \right] N
- t_i^\dagger\left[ i\partial_0 + {\nabla^2\over 4 M}-\Delta_t+\sum_{n=2}^{\infty}c^{(n)}_t\left(i\partial_0 
+ {\nabla^2\over 4 M}\right)^n\right] t^i
\nonumber\\
&&
-s_a^\dagger\left[ i\partial_0 + {\nabla^2\over 4 M}-\Delta_s+\sum_{n=2}^{\infty}c^{(n)}_s\left(i\partial_0 
+ {\nabla^2\over 4 M}\right)^n\right] s^a
\nonumber\\
&&
-y_t \left[ t_i^\dagger N^T P_t^i N+ {\rm h.c.}\right]
- y_s \left[ s_a^\dagger N^T P_s^a N+ {\rm h.c.}\right],
\label{eq:L-dibaryon-0}
\end{eqnarray}
where the isotriplet and isosinglet projectors are defined as
\begin{eqnarray}
P_s^a=\frac{1}{\sqrt{8}}\tau^2\tau^a \otimes \sigma^2, \qquad P_t^i=\frac{1}{\sqrt{8}}\tau^2 \otimes \sigma^2\sigma^i, 
\label{eq:projectors}
\end{eqnarray}
respectively. 
The fully-dressed dibaryon propagators are closely related to the $\si$ ($\siii$) 
scattering amplitudes through
\begin{eqnarray}
i\mathcal{M}_{s(t)}=\frac{4\pi}{M}\frac{i}{k^*\cot\delta_{s(t)}-ik^*}=\frac{y_{s(t)}^2}{-\mathcal{D}_{s(t)}^{-1}+I_0^{ss(tt)}},
\label{eq:scatt-amplitude}
\end{eqnarray}
providing the conditions to match the low-energy constants (LECs) of the Lagrangian of Eq.~\eqref{eq:L-dibaryon-0} at a given renormalization scale, $\mu$, to the low-energy 
scattering parameters. Here $k^*=\sqrt{ME-\bm{P}^2/4}$ is the magnitude 
of the momentum of each nucleon in the center-of-mass frame, $M$ is the nucleon mass, and 
$E$ and $\bm{P}$ are the total energy and momentum of the system, respectively. $\delta_{s(t)}$ 
is the {\it s}-wave phase shift in the isotriplet (isosinglet) channel, and the bare dibaryon propagators are
\begin{eqnarray}
\mathcal{D}_{s(t)}=\frac{-i}{E-\frac{\bm{P}^2}{4M}-\Delta_{s(t)}+\sum\limits_{n=2}^{\infty}c^{(n)}_{s(t)}
\left(E - {\bm{P}^2\over 4 M}\right)^n+i\epsilon}.
\label{eq:bare-prop}
\end{eqnarray}
The quantity $I_0^{ss(tt)}$ in Eq.~(\ref{eq:scatt-amplitude}) is the {\it s}-channel two-nucleon loop diagram that evaluates to
\begin{eqnarray}
I^{ss(tt)}_0=\frac{iM}{4\pi}y_{s(t)}^2(\mu+ik^*)
\label{eq:I0}
\end{eqnarray}
in the power-divergence subtraction scheme~\cite{Kaplan:1998sz,Kaplan:1998we}. 
At momenta below the {\it t}-channel cut, where an effective-range expansion of the scattering amplitude is valid, 
the $s$-wave scattering phase shift can be written in terms of the scattering length
$a_{s(t)}$, the effective range $r_{s(t)}$, and the shape parameters $\rho^{(n)}_{s(t)}$, 
\begin{eqnarray}
k^*\cot\delta_{s(t)}=-\frac{1}{a_{s(t)}}+\frac{1}{2}r_{s(t)}{k^*}^2+\sum_{n=2}^\infty\frac{\rho^{(n)}_{s(t)}}{n!}~(k^{*2})^{n}.
\end{eqnarray}
This leads to matching relations between the LECs of the dibaryon formalism and the low-energy scattering parameters:
\begin{eqnarray}
y_{s(t)}^2=\frac{8\pi}{M^2r_{s(t)}},
\qquad
\Delta_{s(t)}=\frac{2}{Mr_{s(t)}}\left(\frac{1}{a_{s(t)}}-\mu\right),
\qquad 
c^{(n)}_{s(t)}=\frac{2}{Mr_{s(t)}}\frac{\rho^{(n)}_{s(t)}M^n}{n!}.
\end{eqnarray}
%

\subsection{The pionless EFT in background axial fields}
An interaction Lagrangian encoding axial transitions between two-nucleon channels can be constructed out of 
nucleon and dibaryon fields, as well as the background axial field $W^a_i$, where $a$ ($i$) denotes the isovector (vector) indices of the field as before. At leading order (LO) in the EFT, such interactions are momentum independent, and at first order in the background field~\cite{Butler:1999sv, Butler:2000zp, Kong:2000px},\footnote{Since the background field is of arbitrary strength, it 
is not assigned an order in the EFT($\pislash$) power counting. The order of the EFT therefore refers to the low-momentum (derivative) expansion of the interaction terms.
}
\begin{eqnarray}
{\cal L}^{(1)} & \supseteq & -\frac{g_A}{2} N^\dagger \sigma_3 \left[ W_3^- \tau^++W^3_3 \tau^3+W_3^+ \tau^- \right] N
\nonumber\\
&&- \frac{l_{1,A}} {2M\sqrt{r_s r_t}} \left[ W_3^- t_3^\dagger s^++W^3_3t_3^\dagger s^3+W_3^+ t_3^\dagger s^- +\text{h.c.} \right],
\label{eq:L-dibaryon-1}
\end{eqnarray}
where, for simplicity, the background axial field is defined to be nonvanishing only for the $i=3$ component, and $W_\mu^{\pm} \equiv (W_\mu^1\pm i W_\mu^2)/\sqrt{2}$. As will become apparent in Sec.~\ref{sec:M-nnpp-EFT}, it is useful to define a new coupling, $\tilde{l}_{1,A}$, that encapsulates solely two-body contributions to the amplitudes,
\begin{eqnarray}
\tilde{l}_{1,A}=l_{1,A}+2M\sqrt{r_s r_t}g_A.
\label{eq:l1Atilde}
\end{eqnarray}

\begin{figure}[!t]
\centering
\includegraphics[width=0.75\columnwidth]{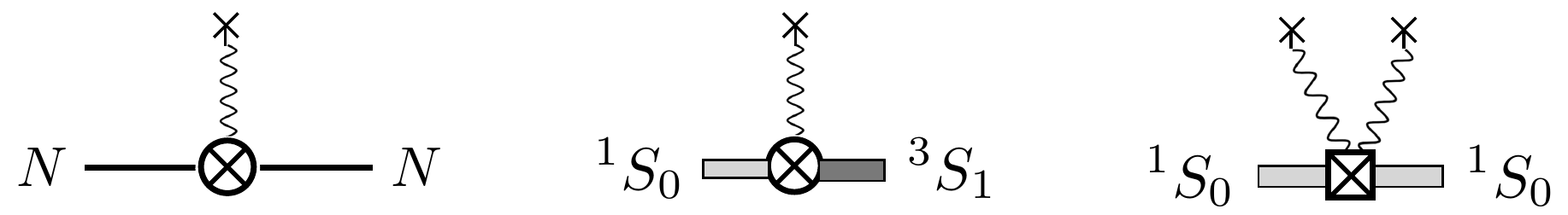}
\caption{
The one-body (left) and two-body (center) operators corresponding to a single insertion of the axial current, $W_\mu^a$, 
described by Eq.~(\ref{eq:L-dibaryon-1}), with coefficients $g_A$ and $l_{1,A}$, respectively.
The 
 two-body operator corresponding to two insertions of the background axial field (right)
described by Eq.~(\ref{eq:L-dibaryon-2}), with coefficient $h_{2,S}$.
The solid, wavy, light-gray and dark-gray thick lines correspond to nucleon fields, axial background fields, and isotriplet and isosinglet dibaryon fields, respectively. 
}
\label{fig:EffectiveOps}
\end{figure}

At second order in the background axial field, multiple terms arise at LO in the expansion, 
including both the single-nucleon and dibaryon interactions with the fields. 
For the $nn$ to $pp$ isotensor transition, the only contribution arises from coupling to an $I=2,~I_3=2$ background field,
\begin{eqnarray}
{\cal L}^{(2)} & \supseteq & - \frac{h_{2,S}} {2Mr_s} {\cW}^{ab} {s^a}^{\dagger} s^b
\supset 
- \frac{h_{2,S}} {2Mr_s} (W_3^+)^2 s^{+\dagger} s^-,
\label{eq:L-dibaryon-2}
\end{eqnarray}
where ${\cW}^{ab}=W^{\{a}_3W^{b\}}_3$ is the symmetric traceless combination of two background fields at the same location. Similar to the $\tilde{l}_{1,A}$ coupling, a new coupling $\tilde{h}_{2,S}$ can be defined to exclude the one-body contributions to the transition amplitudes from the interaction in Eq.~(\ref{eq:L-dibaryon-2}),
\begin{eqnarray}
\tilde{h}_{2,S}=h_{2,S}-\frac{M^2r_s}{2\gamma_s^2}g_A^2.
\label{eq:h2Stilde}
\end{eqnarray}
The three types of interactions with the axial field are shown graphically in Fig.~\ref{fig:EffectiveOps}. For non-maximal isospin transitions, additional operators are needed in Eq.~(\ref{eq:L-dibaryon-2}), but these are not required for the $\beta\beta$-decay process. 

An important aspect of $\beta\beta$ decay is highlighted by Eq.~(\ref{eq:L-dibaryon-2}).
Precise measurements or calculations of single-$\beta$ decay rates in nuclei, including a detailed understanding of the phenomenological quenching of $g_A$ in nuclei, 
are insufficient for high-precision calculations of $\beta\beta$-decay rates.
There are contributions to the matrix element of two axial currents from short-distance physics above the cutoff scale of EFT($\pislash$). These are encapsulated by local operators that do not contribute to single $\beta$-decay matrix elements, but do contribute to $\beta\beta$-decay matrix elements. These are analogues of the two-nucleon electromagnetic polarizability operators, see e.g., Ref.~\cite{Beane:2000fi}. In pionful EFT, the isotensor axial polarizability of a pion exchanged between two nucleons has been argued to provide a dominant contribution to the $nn \to pp$ matrix element~\cite{Prezeau:2003xn} through chiral power-counting of the nucleon-nucleon potential. However, as mentioned previously, this counting is known to be inconsistent in this channel. This contribution is integrated out in EFT($\pislash$), and is therefore encapsulated in the short-distance two-nucleon operator in Eq.~(\ref{eq:L-dibaryon-2}). In addition to  SM effects, such contributions may also be induced in a variety of BSM scenarios~\cite{Prezeau:2003xn, Nicholson:2016byl}. A careful analysis of both  contributions and their mixing will be required in future studies.

\subsection{Correlation functions for the $nn\to pp$ process within pionless EFT
\label{sec:M-nnpp-EFT}}
\begin{figure}[!t]
	\includegraphics[width=1\columnwidth]{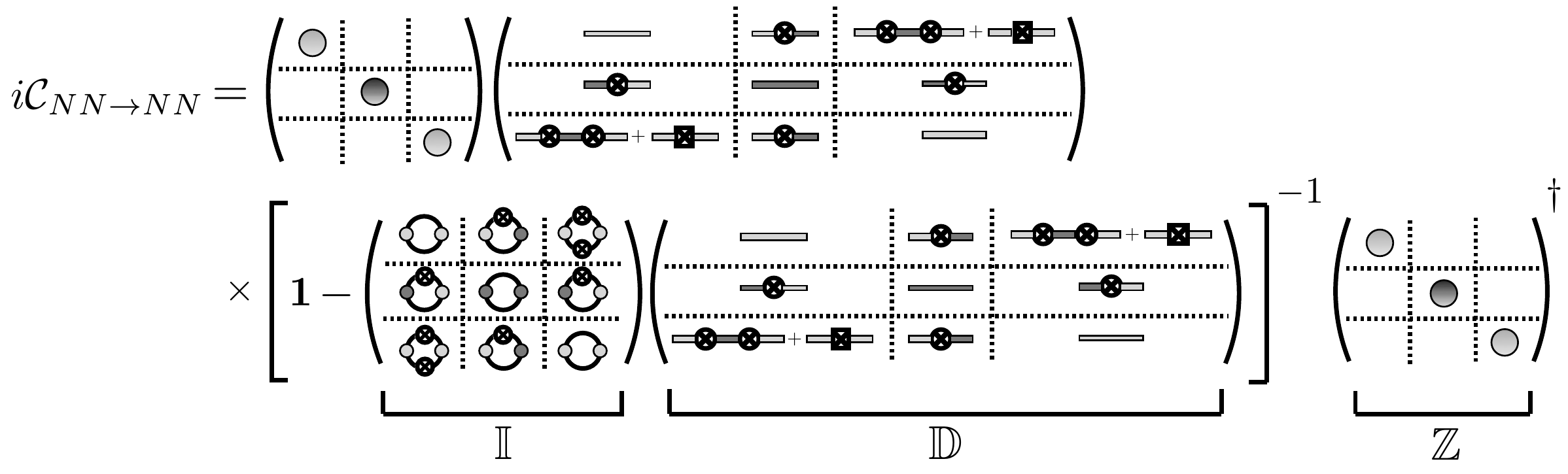}
	\caption{
	Diagrammatic representation of the EFT correlation function matrix in the $\{ nn, np({^3}S_1),pp \}$ coupled-channel space 
	in the presence of a background axial field coupled to the nucleon and dibaryon fields through the interactions displayed in Fig.~\ref{fig:EffectiveOps}.	
	The geometric sum should be expanded to second order in the weak field, and the second-order responses of the diagonal elements of the generalized propagator matrix $\mathbb{D}$ have not been included 
	as they do not affect the $nn \to pp$ transition amplitude to this order. 
	The large light (dark) gray circles denote the overlap function to the isotriplet (isosinglet) dibaryon field. 
	The small light (dark) gray circles denote the isotriplet (isosinglet) dibaryon strong coupling to two nucleons, $y_s$ ($y_t$), 
	while the thick light (dark) gray lines denote the bare isotriplet (isosinglet) dibaryon propagator, $\mathcal{D}_s$ ($\mathcal{D}_t$). 
	The thin black lines are nucleon propagators. 
	The crossed circle denotes the singly weak single-nucleon coupling to the background field when inserted on the 
	nucleon line (proportional to $g_A$), and the singly weak dibaryon coupling when inserted on the dibaryon line 
	(proportional to $l_{1,A}$). 
	Finally, the crossed square represents the doubly weak dibaryon coupling to the background field (proportional to $h_{2,S}$).
	}
	\label{fig:scatt-amplitude}
\end{figure}

\begin{figure}[!t]
	\includegraphics[width=0.95\columnwidth]{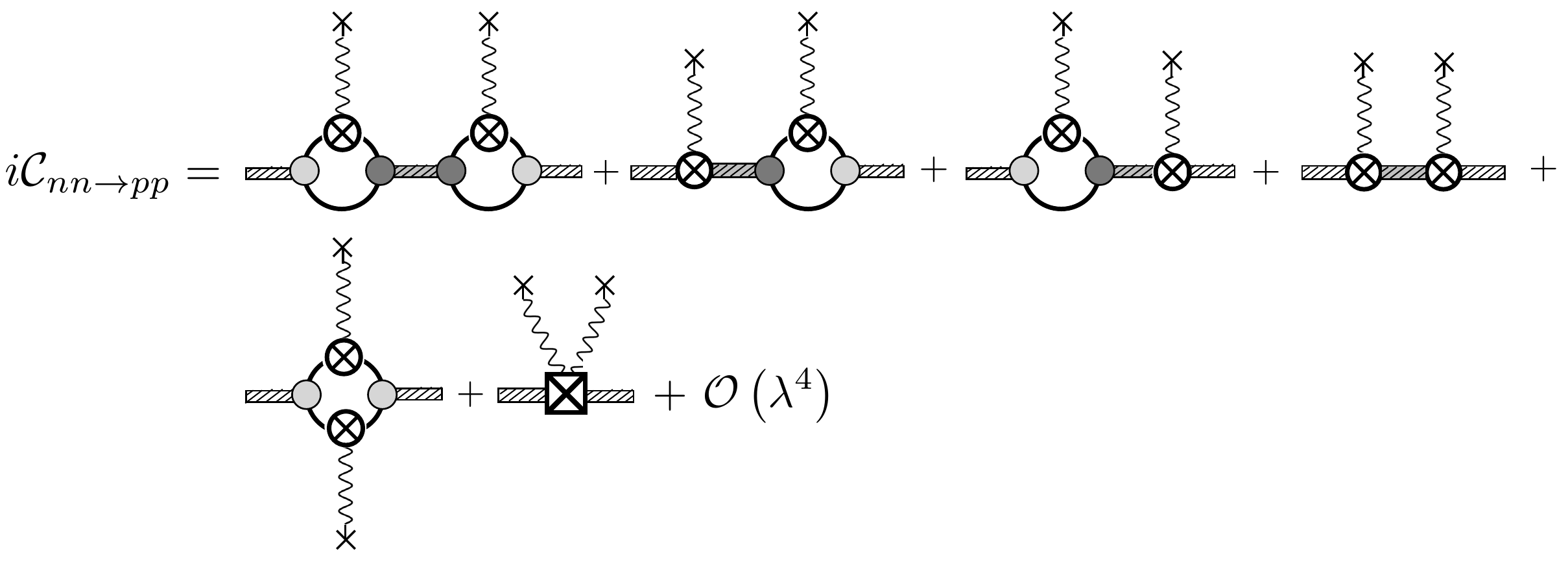}
	\caption{
	Diagrammatic representation of the (unamputated) correlation function for the $nn \to pp$ transition at second order in the axial field, Eq.~(\ref{eq:C-EFT-E-IV}). The small light (dark) gray circles denote the isotriplet (isosinglet) strong dibaryon coupling to two nucleons, $y_s$ ($y_t$), 
	while the thick dashed light (dark) gray lines denote the fully-dressed (by s-wave strong interactions) isotriplet (isosinglet) dibaryon propagator. 
	The thin black lines are nucleon propagators. 
	The crossed circle denotes the singly weak single-nucleon coupling to the background field when inserted on the 
	nucleon line (proportional to $g_A$), and the singly weak dibaryon coupling when inserted on the dibaryon line 
	(proportional to $l_{1,A}$). 
	Finally, the crossed square represents the doubly weak dibaryon coupling to the background field (proportional to $h_{2,S}$). The overlap factors in Eq.~(\ref{eq:C-EFT-E-IV}) are set to unity for simplicity.
	}
	\label{fig:twoax}
\end{figure}

The LECs of the effective Lagrangian, including couplings to the background fields, can be determined by matching correlation functions constructed in the EFT to those computed in LQCD. To study the $nn \to pp$ matrix element induced by the background axial field, it is convenient to construct the correlation function matrix in the $\{ nn, np({^3}S_1),pp \}$ channel space:
\begin{eqnarray}
\mathcal{C}_{NN \to NN} \equiv 
\left(
\begin{array}{ccc}
\cC_{nn \to nn} & \cC_{nn \to np({^3S_1})} & \cC_{nn \to pp} \\ 
\cC_{np({^3S_1}) \to nn} & \cC_{np({^3S_1}) \to np({^3S_1}) } & \cC_{np({^3S_1}) \to pp} \\ 
\cC_{pp \to nn} & \cC_{pp \to np({^3S_1})} & \cC_{pp \to pp} \\ 
\end{array}
\right).
\label{eq:Cmat}
\end{eqnarray}
Note that since the axial background field changes both the spin and isospin, the $np(\si)$ two-nucleon state does not couple to the channels considered in Eq.~(\ref{eq:Cmat}). The elements of this correlation matrix can be expressed in terms of the LECs, 
including couplings to the background axial field. This can be accomplished with the aid of a diagrammatic representation of the correlation function matrix, 
depicted in Fig.~\ref{fig:scatt-amplitude}. 
In momentum space, the expansion can be cast in the following form:
\begin{eqnarray}
i\mathcal{C}_{NN \to NN}(E) = \mathbb{Z} \cdot \mathbb{D}(E) \cdot \frac{1}{\mathbf{1}-\mathbb{I}(E) \cdot \mathbb{D}(E)} 
\cdot \mathbb{Z}^{\dagger},
\label{eq:M-relation}
\end{eqnarray}
where $E$ denotes the total energy of the two-nucleon state and the total three-momentum is projected to zero. The overlap matrix $\mathbb{Z}$ is defined as
\begin{eqnarray}
\mathbb{Z} \equiv \left(
\begin{array}{ccc}
\mathcal{Z}_s & 0 & 0 \\
0 & \mathcal{Z}_t & 0 \\
0 & 0 & \mathcal{Z}_s \\
\end{array}
\right),
\end{eqnarray}
where $\mathcal{Z}_s$ and $\mathcal{Z}_t$ denote the overlaps of interpolating fields onto the isotriplet and isosinglet dibaryon states, respectively. The generalized propagator matrix, $\mathbb{D}$, is defined at second order in the weak field;
\begin{eqnarray}
\mathbb{D} \equiv 
\left(
\begin{array}{ccc}
\mathcal{D}_s & -il'_{1,A}\mathcal{D}_s\mathcal{D}_t\lambda & (-ih'_{2,S}-{l'}_{1,A}^2\mathcal{D}_t){\mathcal{D}_s}^2\lambda^2 \\
-il'_{1,A}\mathcal{D}_s\mathcal{D}_t\lambda & \mathcal{D}_t & -il'_{1,A}\mathcal{D}_s\mathcal{D}_t\lambda \\
(-ih'_{2,S}-{l'}_{1,A}^2\mathcal{D}_t){\mathcal{D}_s}^2\lambda^2 & -il'_{1,A}\mathcal{D}_s\mathcal{D}_t\lambda & \mathcal{D}_s \\
\end{array}
\right),
\end{eqnarray}
to incorporate the effect of channel-changing background field contact interactions on the bare dibaryon propagators. The LECs have been redefined as 
 $l'_{1,A}=\frac{1}{2M\sqrt{r_sr_t}}l_{1,A}$ and $h'_{2,S}=\frac{1}{2Mr_s}h_{2,S}$, 
 and $\lambda~(=W_3^+)$ denotes the strength of the background axial field. 
The matrix of loop functions $\mathbb{I}$ is defined as
\begin{eqnarray}
\mathbb{I} \equiv 
\left(
\begin{array}{ccc}
I_0^{ss} & I^{st}_1\lambda & I^{ss}_2\lambda^2 \\
I^{st}_1\lambda & I^{tt}_0 & I^{st}_1\lambda \\
I^{ss}_2\lambda^2 & I^{st}_1\lambda & I^{ss}_0 \\
\end{array}
\right),
\end{eqnarray}
where $I^{ss(tt)}_0$, $I^{st}_1$ and $I^{ss}_2$ are the $s$-channel two-nucleon loops with zero, 
one and two insertions of the axial field on the nucleon lines, respectively, and with 
appropriate insertions of the strong couplings $y_{s(t)}$ on either side, as shown in Figs. \ref{fig:scatt-amplitude} and \ref{fig:twoax}. 
Note that $I_2^{ss}$ involves single couplings to the axial background field on each of the nucleon propagators 
but no double couplings on a single propagator as the matrix element of an isotensor current between single-nucleon states vanishes.
The value of $I_0^{ss(tt)}$ is given in Eq.~(\ref{eq:I0}), and $I_1^{st}$ and $I_2^{ss}$ are
\begin{eqnarray}
\label{eq:I1}
I^{st}_1=g_A y_sy_t \frac{M^2}{8 \pi k^*}, \ \ \ 
I^{ss}_2=g_A^2 y_s^2 \frac{M^3}{32 \pi {k^*}^3},
\label{eq:I2}
\end{eqnarray}
for $k^{*2}<0$. 
These terms arise from finite loop integrations and do not introduce any further scale dependence. 
The $\mathcal{C}_{NN \to NN}$ matrix elements, Eq.~(\ref{eq:Cmat}),  therefore evaluate to
\begin{eqnarray}
&&i\cC_{nn \to nn} = \frac{\mathcal{Z}_s^2}{\mathcal{D}_s^{-1}-I^{ss}_0}+~\mathcal{O}\left(\lambda^2\right),
\\ 
&&i\cC_{np({^3S_1}) \to np({^3S_1})} = \frac{\mathcal{Z}_t^2}{\mathcal{D}_t^{-1}-I^{tt}_0}+~\mathcal{O}\left(\lambda^2\right),
\label{eq:C-EFT-E-II}
\end{eqnarray}
\begin{eqnarray}
&&i\cC_{nn \to np({^3S_1})}=\frac{ \mathcal{Z}_s\mathcal{Z}_t}{(\mathcal{D}_s^{-1}-I^{ss}_0)(\mathcal{D}_t^{-1}-I^{tt}_0)}(I^{st}_1-il'_{1,A}) \lambda+~\mathcal{O}\left(\lambda^3\right),
\label{eq:C-EFT-E-III}
\\
&&i\cC_{nn \to pp} =\frac{\mathcal{Z}_s^2}{\mathcal{D}_s^{-1}-I^{ss}_0}\left[\frac{I^{ss}_2-ih'_{2,S}}{\mathcal{D}_s^{-1}-I^{ss}_0}
+
\frac{(I^{st}_1-i{l'}_{1,A})^2}{(\mathcal{D}_s^{-1}-I^{ss}_0)(\mathcal{D}_t^{-1}-I^{tt}_0)} \right] \lambda^2+~\mathcal{O}\left(\lambda^4\right),
\label{eq:C-EFT-E-IV}
\end{eqnarray}
where the energy dependence of the functions has been suppressed. Diagrams representing the various contributions to $\cC_{nn \to pp}$ are shown in Fig.~\ref{fig:twoax}.

%

\subsection{Matching to LQCD correlation functions}
To match to the analogous LQCD correlation functions, the finite-volume counterpart of Eq.~(\ref{eq:C-EFT-E-IV}) must be used
with periodic boundary conditions in a cubic spatial volume. 
Furthermore, the energy-dependent correlation function must be Fourier transformed in time and then rotated to Euclidean space, 
i.e. $x_0 \to i t$. 
The only finite-volume effects that are not exponentially suppressed below the two-particle inelastic thresholds arise when intermediate two-nucleon 
states can be on their mass shell. This can only happen within the $s$-channel loops. 
In these loops, the integration is replaced by a summation over quantized momenta, and the singularities of the summand, 
corresponding to the on-shell condition, give rise to either power-law volume corrections for scattering states or 
exponential corrections for bound states. All other quantities in Eq.~(\ref{eq:C-EFT-E-IV}), including the bare dibaryon 
propagators and the overlap functions, are equivalent to their infinite-volume counterparts up to exponential corrections 
that are suppressed with the range of nuclear forces (set by the pion Compton wavelength). 
The $s$-channel loops in a finite volume, denoted as $\mathcal{I}$ below, can be evaluated straightforwardly, but their forms are not needed in this work as will be discussed below. The main finite-volume characteristic of the correlation functions that must be accounted for is the discrete nature of the two-particle finite-volume spectra, arising from the quantization conditions~\cite{Luscher:1986pf, Luscher:1990ux, Beane:2003da, Briceno:2013lba}:
\begin{eqnarray}
\left . (\mathcal{D}_{s(t)}(E)^{-1}-\mathcal{I}^{ss(tt)}_0(E)) \right |_{E=E^{(n)}_{s(t)}}=0,
\label{eq:QC}
\end{eqnarray}
where $E^{(n)}_{s(t)}$ are the discrete finite-volume energy eigenvalues of the two-nucleon isotriplet (isosinglet) channels in the absence of the background axial field. Here, the effects of the nonzero lattice spacing and finite temporal extent are ignored. As a result, the Fourier transform of the correlation functions can be obtained by performing an integration over a continuous energy variable. This integration is straightforward, given the known  energy dependence of the correlation functions, shown in Eq.~(\ref{eq:QC}). One subtlety is an apparent singular behavior of the loop functions $I_1^{st}$ and $I_2^{ss}$ at $E=0$, which naively introduces further contributions to the energy integral. These singularities are an artifact of the finite-order expansion of the correlation function in the weak fields. A straightforward exercise shows that the all-order correlation function in Eq.~(\ref{eq:C-EFT-E-IV}) does not contain such singularities. Therefore, this correlation function must be first Fourier transformed in time and then expanded in the weak field. The result of this procedure is identical to Fourier transforming the second-order correlation function in Eq.~(\ref{eq:C-EFT-E-IV}) as long as such spurious singularities are neglected.  

To obtain the matrix elements, it is necessary to take the ratio of the $nn \to pp$ three-point function in the background field,
\begin{eqnarray}
&& C_{nn \to pp} (t)= 
\lambda^2
\sum_{n} e^{-E^{(n)}_s t} 
{\mathcal{Z}_s^{(n)}}^2 
\mathcal{R}_s(E^{(n)}_s) 
\left[
\vphantom{\frac{\mathcal{R}_s(E^{(n)}_s)\mathcal{R}_t(E^{(l)}_t)
 \left(\mathcal{I}^{st}_1(E^{(n)}_s)-il'_{1,A}\right)^2}{E^{(n)}_s-E^{(l)}_t} }\ 
t\ \mathcal{R}_s(E^{(n)}_s) 
\left(\mathcal{I}_2^{ss}(E^{(n)}_s)-ih'_{2,S}\right)
\right.\nonumber\\
&& \left.
\qquad
\ +\ 
 \sum_{l}\frac{\mathcal{R}_s(E^{(n)}_s)\mathcal{R}_t(E^{(l)}_t)
 \left(\mathcal{I}^{st}_1(E^{(n)}_s)-il'_{1,A}\right)^2}{E^{(n)}_s-E^{(l)}_t} 
 \left(t - \frac{e^{(E_s^{(n)}-E^{(l)}_t) t}-1}{E_s^{(n)}-E^{(l)}_t} \right)
 \ +\ \dots
\right],
\label{eq:C-EFT-t-IV}
\end{eqnarray}
to the zero-field two-point function,
\begin{eqnarray}
&&
C_{nn \to nn}(t) =
-\sum_{n} 
e^{-E^{(n)}_s t} \ 
{\mathcal{Z}_s^{(n)}}^2 \ 
\mathcal{R}_s(E^{(n)}_s),
\label{eq:C-EFT-t-I}
\end{eqnarray}
where $\mathcal{R}_{s(t)}$ is related to the residue of the fully-dressed dibaryon propagator evaluated at the finite-volume energies $E^{(n)}_{s(t)}$,
\begin{eqnarray}
\mathcal{R}_{s(t)}(E^{(n)}_{s(t)})=\left[ \left . \frac{d}{dE}(\mathcal{D}_{s(t)}(E)^{-1}-\mathcal{I}^{ss(tt)}_0(E)) \right|_{E=E^{(n)}_{s(t)}} \right]^{-1}.
\label{eq:Res}
\end{eqnarray}
In Eq.~(\ref{eq:C-EFT-t-IV}) and (\ref{eq:C-EFT-t-I}), $\mathcal{Z}_s^{(n)}$ is the overlap of the interpolating fields onto the states of quantized energy.
The ellipsis in Eq.~(\ref{eq:C-EFT-t-IV}) denotes additional terms that are time independent, or have a time dependence that is exponentially suppressed by the energy gaps to the excited states, which are assumed to be large in this analysis. Not all terms with time dependence $e^{\Delta t}$ are made explicit in Eq.~(\ref{eq:C-EFT-t-IV}). Among such terms are those that involve transition matrix elements to excited states. These are analogous to the $D$ terms in the LQCD correlation functions analyzed in Eq.~(\ref{eq:funcform}) and are irrelevant to the discussion of the ground-state to ground-state matrix elements. Note that the summations over intermediate states in the EFT context are  over finite-volume scattering states that are explicit degrees of freedom in the EFT, that is, those states with momenta below the cutoff $\Lambda\sim m_\pi$. This should be contrasted with the sums over intermediate states in Sec. \ref{sec:LQCD}, where the states are the eigenstates of (L)QCD. Part of the latter summation is incorporated in the short-distance EFT couplings through the matching, with the cutoff scale defining the separation.

Taking the ratio of Eq.~(\ref{eq:C-EFT-t-IV}) to two times Eq.~(\ref{eq:C-EFT-t-I}), as done for the ratio of LQCD correlation functions in Eq.~(\ref{eq:Rnnpp}), and taking the second derivate of the correlation function with respect to the background-field strength, the second-order finite-volume matrix element can be obtained from the terms linear in time, giving
\begin{eqnarray}
M^{(V)}_{nn \to pp}
&=& 
-
 \mathcal{R}_s(E^{(0)}_s) \ 
\left[
\vphantom{\ -\ 
\left( {\cal I}^{st}_1(E^{(0)}_s)-il'_{1,A} \right)^2
\sum_{l\ne d} {\mathcal{R}_t(E_t^{(l)})  \over \delta^{(l)}}
}
\mathcal{R}_t(E^{(0)}_t) { ({\cal I}^{st}_1(E^{(0)}_s)-il'_{1,A})^2 \over\Delta}
\ +\ {\cal I}_2^{ss}(E^{(0)}_s)
- ih'_{2,S}
\right.\nonumber\\
&&\left. \qquad\qquad\qquad\qquad
\ -\ 
\left( {\cal I}^{st}_1(E^{(0)}_s)-il'_{1,A} \right)^2
\sum_{l\ne d} {\mathcal{R}_t(E_t^{(l)})  \over \delta^{(l)}}
\right]
\ +\ \dots,
\label{eq:M-nnpp-EFT}
\end{eqnarray}
where the first term is
the contribution from the deuteron intermediate state, 
and where the ellipsis denotes terms that are higher order in the EFT($\pislash$) expansion. The remaining short-distance contributions are constrained by matching to LQCD correlation functions. This can be most cleanly demonstrated by defining a new quantity that encapsulates all of the short-distance contributions, including those arising from intermediate states other than the deuteron,
\begin{eqnarray}
\overline{h}^{(V)}_{2,S} 
=
 h_{2,S}
 \ -\ 
 2iM r_s
 \left( {\cal I}^{st}_1(E_s^{(0)})-i\tilde{l}_{1,A} \right)^2
 \sum_{l \neq d}\frac{\mathcal{R}_t(E^{(l)}_t)}{\delta^{(l)}},
\end{eqnarray}
where the superscript denotes that this quantity is volume dependent, with a well-defined infinite-volume limit, $\overline{h}^{(V)}_{2,S} = \overline{h}^{(\infty)}_{2,S} \equiv \overline{h}_{2,S}$. As already discussed in Sec.~\ref{subsec:FV}, the initial and final states, as well as the propagating intermediate state, are deeply bound two-nucleon states in the calculations performed in this work, resulting in exponentially suppressed volume corrections. The infinite-volume limit of all of the contributions to 
$M^{(V)}_{nn \to pp}$ in Eq.~(\ref{eq:M-nnpp-EFT}) can then be taken and, up to a sub-percent uncertainty from volume effects, the infinite-volume matrix element is obtained,
\begin{eqnarray}
M_{nn \to pp}
& = &  
-\frac{|M_{pp \to d}|^2}{\Delta} 
\ +\ 
\frac{Mg_A^2}{4 \gamma_s^{2}} 
-\mathbb{H}_{2,S},
\label{eq:M-nnpp-EFTb}
\end{eqnarray}
where $\mathbb{H}_{2,S}=\frac{\gamma_sZ_s^2}{2M}(\overline{h}_{2,S}-\frac{M^2r_s}{2\gamma_s^2}g_A^2)$ encapsulates the correlated two-nucleon two-axial coupling contribution to the amplitude, and
\begin{eqnarray}
M_{pp \to d}
& = & g_A(1+S)+\mathbb{L}_{1,A}
\label{eq:M-ppd-EFT}
\end{eqnarray}
is the EFT matrix element for the $pp \to d$ process. Here $\mathbb{L}_{1,A}=\frac{Z_s Z_t
\sqrt{\gamma_t \gamma_s}}{2 M}\tilde{l}_{1,A}$ denotes the correlated two-nucleon axial contribution to $M_{pp \to d}$. In the two/few-body sector, this is equivalent to the phenomenological quenching of $g_A$. In Eq.~(\ref{eq:M-ppd-EFT}), $S=-1+Z_s Z_t(\sqrt{\gamma_t/\gamma_s}-\sqrt{\gamma_sr_s}\sqrt{\gamma_tr_t})$ is an $SU(4)$ Wigner symmetry-breaking factor.
In these equations, $\mathcal{R}_{s(t)}(E^{(0)}_{s(t)})=i\gamma_{s(t)} r_{s(t)} Z_{s(t)}^2$, 
with $\gamma_{s(t)}=\sqrt{MB_{s(t)}}$ (where $B_{s(t)}=2M-E^{(0)}_{s(t)}$ is the binding energy) and $Z_{s(t)}^2=1/(1-\gamma_{s(t)} r_{s(t)})$, and the tower of shape parameters has been ignored. The first term in Eq.~(\ref{eq:M-nnpp-EFTb}) corresponds to the deuteron pole, while the second and third terms are short-distance contributions.

The quantities $\mathbb{L}_{1,A}$ and $\mathbb{H}_{2,S}$ can, in principle, be constrained from the values of the proton axial charge and the matrix elements for the $pp \to d$ and $nn \to pp$ processes extracted in Sec.~(\ref{sec:results}). Additionally, the $SU(3)$ flavor-symmetric values of binding momenta and effective ranges are needed, which have been determined in Refs.~\cite{Beane:2012vq,Beane:2013br}. Unfortunately, given the modest $\mathcal{O}\left( 10 \% \right)$ uncertainties on these parameters, the dinucleon and deuteron wavefunction renormalization factors, $Z_s$ and $Z_t$, do not have well-behaved statistical distributions, leading to a broad distribution of the $SU(4)$-breaking function $S$ in Eq.~(\ref{eq:M-ppd-EFT}). As a result, no significant bound can be put on the value of $\mathbb{L}_{1,A}$.\footnote{A variant of this coupling with $S=0$ is defined in Ref.~\cite{Savage:2016kon} as $-L^{2b,sd}_{1,A}$. Since the values of the relevant binding momenta and effective ranges are known much more precisely at the physical point, these values were used in that work to definitively constrain the physical $pp$-fusion matrix element, assuming a mild quark-mass dependence for the correlated two-nucleon axial coupling.} However, $\mathbb{H}_{2,S}$, which is the main focus of this section, is independent of the values of $Z_s$ and $Z_t$, as is evident from  Eq.~(\ref{eq:M-nnpp-EFTb}). This quantity can thus be cleanly extracted:
\begin{eqnarray}
\mathbb{H}_{2,S}=4.7(1.3)(1.8)~\tt{fm},
\label{eq:H2S-vale}
\end{eqnarray}
where the first uncertainty is the combined statistical and systematic uncertainty of a chosen analysis and the second uncertainty covers the differences between the values obtained from different analyses. Although being significantly smaller than the dominant deuteron-pole term, this term is of the same order of magnitude as the second term in the right-hand side of Eq.~(\ref{eq:M-nnpp-EFTb}) (the term proportional to $g_A^2$), and is non-negligible.
The difference between the full matrix element for $nn \to pp$ and the Born term is roughly $5\%$ of the total. Interestingly, this is comparable to the contribution of $l_{1,A}$ to the matrix element.

Once the LECs of EFT($\pislash$) for both the first-order $\Delta I=1$ and the second-order $\Delta I=2$ interactions with an axial background field are determined, they can be used in few-body calculations to make predictions for the $\beta\beta$-decay matrix elements in light nuclei
at the quark masses used in this LQCD study. 
An example of such an approach for spectroscopy is given in Ref.~\cite{Barnea:2013uqa,Contessi:2017rww}, 
and the extension to electroweak interactions is in progress~\cite{BarneaPC}. 
Eq.~(\ref{eq:H2S-vale}) is only valid at  the heavy quark masses that are used in the LQCD calculations, and to connect directly to phenomenology the physical quark masses must be used. 
Alternatively, using unphysical quark masses that are sufficiently close to the physical values, 
pionful EFT could be used to make phenomenological predictions via extrapolations. 
In either situation, the relation between the finite-volume bi-local matrix elements and the infinite-volume transition 
amplitudes is more complicated due to the scattering nature of states involved, 
and a generalization of the formalism presented in Ref.~\cite{Christ:2015pwa} to address this situation is in progress~\cite{Davoudi:2017}.

\section{Summary and Outlook
\label{sec:sum}}

An observation of nuclear $0\nu\beta\beta$ decay  would provide unambiguous evidence for the 
violation of lepton number and the Majorana nature of neutrinos.
Lepton-number violation can manifest in $0\nu\beta\beta$ decay in distinct ways; for example, in the form of the exchange of a light Majorana neutrino, or through local operators arising from new physics above the electroweak scale, with the most relevant of these being four-quark-two-electron operators.   
Both the  lepton-number conserving $2\nu\beta\beta$-decay modes 
and the lepton-number violating $0\nu\beta\beta$-decay modes induced by a light Majorana neutrino
depend upon nuclear matrix elements with two insertions of the weak currents.\footnote{At the scale of chiral-symmetry breaking, matching lepton-number violation induced by a Majorana neutrino 
	to the low-energy EFT will give rise to effective operators with the same structure as some of the operators induced by 
	four-quark-two-lepton operators originating at high scales.  
	The relative size of the contributions, and the ability of future measurements to distinguish between these origins, remain to be explored.}   
At the scale of the strong interactions, these receive both long-distance and short-distance contributions.  
The long-distance contributions are largely dictated by the single-nucleon axial matrix element, $g_A$, and by correlated two-nucleon interactions (meson-exchange currents).  
The short-distance contributions, from physics above the chiral symmetry breaking scale, 
are encapsulated in the isotensor axial polarizability which does not contribute to single $\beta$-decay rates.
 Such contributions are, furthermore, in addition those induced by a finite nuclear model space.  

In this paper, a detailed investigation of the second-order weak $nn\rightarrow pp$ transition matrix element is 
presented using LQCD and EFT, expanding upon the results presented in Ref.~\cite{Shanahan:2017bgi}.
In particular, the long-distance Born term and the short-distance contributions are explicitly separated in 
LQCD calculations performed at unphysical values of the quark masses corresponding to $m_\pi\sim 806~{\tt MeV}$, at one lattice spacing 
and in one lattice volume.  
The short-distance contribution, in the language of EFT, receives contributions from two-nucleon states involving momenta below the cutoff and from a local operator encapsulating shorter-distance physics.
The LQCD calculations utilize the recently-developed fixed-order background-field approach~\cite{Savage:2016kon} to cleanly isolate matrix elements corresponding to a fixed number of insertions of the isovector axial current.  
Further details of this method, along with the associated analysis techniques used to extract the $nn\rightarrow pp$ transition matrix element, are presented.
Second-order weak processes are discussed in the dibaryon formulation of pionless EFT whose finite-volume Euclidean-space correlation functions
are constructed and matched to the LQCD correlation functions, 
allowing a determination of the leading two-nucleon second-order weak coupling.
In conjunction with many-body methods, these  couplings can be used to predict $\beta\beta$-decay rates of nuclei at these quark masses.
The isotensor axial polarizability is found to provide a non-negligible contribution to the $nn\rightarrow pp$ matrix element. This  contribution will need to be determined at the physical values of quark masses to impact the experimental program.
As the isotensor axial polarizability operators do not contribute to single-$\beta$ decay, using a quenched value of $g_A$ does not account for this physics. 
This is a previously ignored contribution to nuclear $\beta\beta$ decays that can only be constrained experimentally by $\beta\beta$-decay rates, and requires further exploration, in particular using LQCD.

The methods developed in this work have applications beyond the determination of second-order axial responses at threshold. 
The extension of the present study to the case of $0\nu\beta\beta$ decay in the light Majorana scenario involves additional 
challenges arising from the loop integration over intermediate states. 
It is likely that a new approach will be required to address this, for which preliminary work is underway~\cite{Davoudi:2017}. 
Nevertheless, with better constraints on $2\nu\beta\beta$-decay rates, the accuracy of predictions for $0\nu\beta\beta$-decay 
rates is expected to improve.
In addition, the technology developed in this work can be utilized to study second-order responses that are relevant in assessing the effects of two-photon contributions to electromagnetic form factors, and for calculating the $\gamma Z$ box diagram relevant for parity-violating electron-proton scattering.

In the future, the calculations presented in this work will be extended to lighter quark masses, larger lattice volumes and multiple lattice spacings, accounting for the dominant systematics that remain unexplored. 
Significant difficulties are anticipated in taking these steps, in particular given the bi-local nature of the quantities that are considered.
As the quark masses reach their physical values, and the volumes become larger, the hierarchy between the dinucleon-deuteron 
mass splitting, $\Delta$, and the gap to excitations of the dinucleon system, $\delta$, is changed ($\delta\rightarrow 0$ and 
$\Delta\rightarrow 2.22~{\tt MeV}$), thereby making the current analysis strategy ineffective, 
as the contributions that involve the transition matrix elements of excited states will no longer be negligible.
Separating the source and sink timeslices from the region of the background field will ameliorate this problem~\cite{Christ:2012se}, 
but the extraction of the relevant long and short-distance contributions to the $nn\rightarrow pp$ matrix element will remain complicated. 
Nevertheless, we anticipate that LQCD calculations with physical quark masses will provide essential input to many-body calculations of \tnubb\ and \znubb-decay rates that cannot be obtained through any other known method.

\section*{Acknowledgments:}
This research was supported in part by the National Science Foundation under grant number NSF PHY11-25915 and
we acknowledge the Kavli Institute for Theoretical Physics for hospitality 
during preliminary stages of this work.
Calculations were performed using computational resources provided
by NERSC (supported by U.S. Department of
Energy grant number DE-AC02-05CH11231),
and by the USQCD
collaboration. This research used resources of the Oak Ridge Leadership 
Computing Facility at the Oak Ridge National Laboratory, which is supported 
by the Office of Science of the U.S. Department of Energy under Contract 
number DE-AC05-00OR22725. 
The PRACE Research Infrastructure resources at the 
Tr\`es Grand Centre de Calcul and Barcelona Supercomputing Center were also used.
Parts of the calculations used the {\tt chroma} software
suite~\cite{Edwards:2004sx} and the {\tt quda} library~\cite{Clark:2009wm,Babich:2011np}. 
EC was supported in part by the USQCD SciDAC project, the U.S. Department of Energy through 
grant number DE-SC00-10337, and by U.S. Department of Energy grant number DE-FG02-00ER41132.
ZD, WD and PES were partly supported by U.S. Department of Energy Early Career Research Award DE-SC0010495 and grant number DE-SC0011090. The  work of WD is supported in part by the U.S. Department of Energy, Office of Science, Office of Nuclear Physics, within the framework of the TMD Topical Collaboration. 
KO was partially supported by the U.S. Department of Energy through grant
number DE-FG02-04ER41302 and through contract number DE-AC05-06OR23177
under which JSA operates the Thomas Jefferson National Accelerator Facility. 
MJS was supported by DOE grant number~DE-FG02-00ER41132, and in part by the USQCD SciDAC project, 
the U.S. Department of Energy through grant number DE-SC00-10337.	
BCT was supported in part by a joint City College of New York-RIKEN/Brookhaven Research Center
fellowship, and by the U.S. National Science Foundation, under grant
number PHY15-15738. 
MLW was supported in part by DOE grant number~DE-FG02-00ER41132.
FW was partially supported through the USQCD Scientific Discovery through Advanced Computing (SciDAC) project 
funded by U.S. Department of Energy, Office of Science, Offices of Advanced Scientific Computing Research, 
Nuclear Physics and High Energy Physics and by the U.S. Department of Energy, Office of Science, Office of Nuclear Physics under contract DE-AC05-06OR23177.
We gratefully acknowledge compute resources on the capacity computing hardware at Jefferson Lab provided under the Lattice QCD Computing Project Extension II, allocated to this investigation by the USQCD Scientific Program Committee.

\bibliography{bibi.bib}
\end{document}